\PassOptionsToPackage{dvipsnames}{xcolor}
\documentclass[11pt]{article}
\usepackage[margin=1in]{geometry}
\usepackage[numbers]{natbib}

\usepackage{graphicx, amsmath, amsthm, cleveref}
\graphicspath{{./Images}}

\usepackage{amsfonts}
\usepackage{graphicx}
\usepackage{amsmath}
\usepackage{amssymb}

\ifpdf
  \DeclareGraphicsExtensions{.eps,.pdf,.png,.jpg}
\else
  \DeclareGraphicsExtensions{.eps}
\fi

\title{\bfseries Fast Nonparametric Estimation of Sparse Precision Matrices with A Priori Sparsity Structure}

\author{Caleb Derrickson\thanks{Argonne National Laboratory, Lemont, IL.}
\and Alexis Montoison\footnotemark[1]
\and Vishwas Rao\footnotemark[1]
\and Christopher J. Geoga\thanks{University of Wisconsin-Madison, WI (\texttt{geoga@wisc.edu})}\thanks{This work is partially funded by the National Science Foundation under Grant Number DMS-2610202.}
\and Mihai Anitescu\footnotemark[1]
}

\date{}

\usepackage{amsopn}
\DeclareMathOperator{\diag}{diag}

\makeatletter
\newcommand*{\addFileDependency}[1]{
  \typeout{(#1)}
  \@addtofilelist{#1}
  \IfFileExists{#1}{}{\typeout{No file #1.}}
}
\makeatother

\usepackage{multirow}
\usepackage{tikz}

\usepackage{bm}
\let\bmsym\bm
\renewcommand{\bm}[1]{\mathbf{#1}}

\newcommand{\T}{^{\intercal}}
\newcommand{\inv}{^{-1}}

\newcommand{\bx}{\bm{x}}
\newcommand{\by}{\bm{y}}
\newcommand{\bS}{\bm{\Sigma}}
\newcommand{\bOm}{\bm{\Omega}}
\newcommand{\bT}{\bm{T}}
\newcommand{\E}{\mathbb{E}}
\newcommand{\bO}{\mathcal{O}}
\newcommand{\R}{\mathbb{R}}

\begin{document}

\maketitle

\begin{abstract}
This work describes a computational approach for rapidly estimating and
characterizing the covariance structure of a Gaussian random field from a modest
number of i.i.d. replicates by formulating the problem in terms of a sparse
inverse Cholesky factor of the data's covariance matrix. Borrowing from the
extensive literature of using directed acyclic graphs to impose sparsity on
inverse Cholesky factors (for example, Vecchia's approximation or
nearest-neighbor Gaussian process models), a probabilistically parsimonious a
priori sparsity pattern is imposed, and with careful formulation the estimation
problem is given as an extremely well-structured convex problem that can be
solved for millions of unknowns in seconds on a single processor core.  After
describing the problem and specific formulation of the optimization, we
demonstrate the estimator's speed and effectiveness at capturing a variety of
nonstationary structures in spatial or spatio-temporal data, and in ensemble
data assimilation.

\end{abstract}



\section{Introduction} \label{sec:introduction}

A Gaussian process (GP) is a stochastic process $Y(\bx)$ whose
finite-dimensional marginal distributions are multivariate normal distributed.
GPs are entirely specified through their \emph{mean} and \emph{covariance}
functions, which are given by $\mu(\bx) = \E Y(\bx)$ and $K(\bx, \bx') =
\text{Cov}(Y(\bx), Y(\bx'))$. For any finite collection of locations $\{ \bx_j
\}_{j=1}^n$, then, the collection of measurements $\by = [Y(\bx_j)]_{j=1}^n$ has
distribution $\by \sim \mathcal{N}(\bmsym{\mu}, \bS)$, where $\mu_j =
\mu(\bx_j)$ is the mean and $\Sigma_{j,k} = K(\bx_j, \bx_k)$, with $\bS \succeq
\bm{0}$, is the covariance matrix. For the duration of this work, we
will assume that $\mu(\bx) \equiv 0$, which is a standard assumption in the
literature due to the fact that means are often estimated separately and removed
in a pre-processing step.

A fundamental problem in this space is the estimation of $\bS$ from i.i.d.
replicates $\{ \by_s \}_{s=1}^m$. One typically estimates $\bS$ by minimizing
the negative
\emph{log-likelihood}, which is given (with constants suppressed) as
\begin{equation} \label{eq:nll}
  \ell(\bS; \; \by) = \frac{1}{2} \left(
    m \log | \bS | + \sum_{s=1}^m \by_s\T \bS^{-1} \by_s
  \right),
\end{equation}
where $| \bS |$ is the determinant of $\bS$.  Since $\bS$ has $\bO(n^2)$
nonzeros and one has $\bO(n)$ data points, clearly a fully naive estimator for
$\bS$ using, for example, semidefinite programming would be ill-posed. An
extremely strong variety of regularization (although it is not often discussed
in these terms) is a parametric covariance model, which is to assume that the
covariance function $K$ belongs to some parametrically indexed family $\{
K_{\bmsym{\theta}} \}_{\bmsym{\theta} \in \Theta}$, so that
$\Sigma_{j,k} = K_{\bmsym{\theta}}(\bx_j, \bx_k)$ and then to reduce the problem
of estimating $\bS$ to the ubiquitous parametric maximum likelihood problem of
\begin{equation} \label{eq:parametric_nll}
  \hat{\bmsym{\theta}}_{\text{MLE}} =
  \text{argmin}_{\bmsym{\theta} \in \Theta} 
  \ell(\bS(\bmsym{\theta}); \; \by).
\end{equation}
While very strong, the assumption has several extremely attractive qualities:
the optimization problem is reduced to just a few decision variables, the
interpretation of $\bS$ becomes much more direct, and such a descriptive model
that specifies the entire process---as opposed to just estimating the covariance
matrix for one specific finite-dimensional distribution---means that downstream
tasks like prediction at new locations are direct. The downsides of this
approach, however, are that (i) the specific functional form
$K_{\bmsym{\theta}}$ may not describe the true covariances accurately, and (ii)
that the optimization problem of \cref{eq:parametric_nll} can be challengingly
nonconvex, and even with second order information a general-purpose nonconvex
solver may struggle to solve it. From a more purely computational perspective, a
separate category of challenge in this space that neither nonparametric nor the
parametric estimators described here address is the computational burden.
Evaluating \cref{eq:nll} for dense $\bS$ requires a Cholesky factorization of
$\bS$, meaning that the runtime cost of evaluation will grow as $\bO(n^3)$
regardless of how $\bS$ is parameterized.

This work introduces an estimation framework that addresses both of these issues
at once. The key conceptual contribution here is that, in the space of
parametric models, there is a rich literature of models connecting structured
precision matrices $\bOm = \bS^{-1}$ that have a sparsity structure induced by a
directed acyclic graph (DAG) built from the geometry of measurement locations
$\{ \bx_j \}_{j=1}^n$. These approximations go by many names, such as Vecchia's
approximation \cite{vecchia1988,stein2004,katzfuss2021}, nearest-neighbor
Gaussian processes \cite{datta2016,finley2019,jin2024}, conditional
autoregressive (CAR) models \cite{besag1974}, Gauss-Markov random fields
\cite{rue2005}, and more. But the core of each idea is the same: due to the
probabilistic interpretability of precision matrices, structural assumptions
pertaining to conditional independence directly correspond to sparsity structure
in $\bOm$, and even more directly in terms of a triangular factor $\bT$ such
that $\bOm = \bT \bT\T$, as $\bT$ can be interpreted as a transport map whose
columns are scaled linear prediction weights \cite{pourahmadi1999}. By exploiting
this connection and the DAG-based framework for designing sparse approximations
to $\bT$, we will present a specific formulation of the estimator for
$\bT$ with $\bO(nk)$ decision variables, where $k$ is the number of conditioning
points per column, that can be solved extremely efficiently. The estimation
problem is convex in $\bT$; we solve a smooth reparameterization of it whose
stationary points coincide with those of the convex problem, as detailed in
\cref{sec:optdesign}. We will also discuss regularization-type options that
preserve this convexity and have natural probabilistic interpretations,
connecting the nonparametric estimator with well-studied parametric ones.

There is a significant body of work on the nonparametric sparse estimation of
$\bOm = \bS^{-1}$ or $\bT$ such that $\bOm = \bT \bT\T$. The most closely
related method to what we introduce here is given in \cite{kidd2020,boyles2021},
which is equivalent in terms of using Vecchia approximations to obtain an a
priori sparsity structure, but takes a Bayesian approach to estimation. A much
broader category of methods are those that attempt to \emph{discover} the
sparsity of $\bOm$. In \cite{carvalho2009}, for example, Wishart g-priors are
used to discover dependency graphs. Lasso- and Dantzig-like penalties have also
been used in a variety of ways: the graphical lasso \cite{friedman2008} is a
popular and direct application of $\ell_1$ regularization, \cite{cai2011} gives
a similar dual-like form, and \cite{levina2008} introduces a nested lasso-like
penalty that guarantees sparsity in the resulting estimate for $\bOm$. These
sparsity-discovering methods come with strong theoretical guarantees in some
settings, but ultimately have difficult runtime complexity with respect to the
size of the matrix. A particular focus of this estimator is good runtime
complexity and prefactors, so that one can estimate $\bOm$ in $\R^{10^6 \times
10^6}$ and beyond in minutes or less.


\section{Vecchia's approximation} \label{sec:setup}

The parametric covariance estimation problem of solving
\cref{eq:parametric_nll} poses several challenges. First, the negative
log-likelihood is often \emph{highly} nonconvex, and individual parameters in a
parametric model are often not identifiable \cite{zhang2004}. As one might
expect, this produces complex level sets of the likelihood surface that are
highly non-quadratic, and in turn makes navigating them even with a second-order
optimizer challenging. This problem is sidestepped by the nonparametric
formulation \cref{eq:nll} which directly optimizes over $\bS$. As mentioned in
the Introduction, we elect to avoid semidefinite programming by instead
parameterizing $\bS^{-1}$ in terms of its (permuted) Cholesky factor, which
retains convexity and interpretability.  The issue that is not solved by this
nonparametric formulation, however, is scalability. While methods such as the
graphical lasso may result in a sparse approximation to $\bS^{-1}$
such that downstream operations with it can be performed efficiently,
\emph{discovering} the sparsity itself is challenging.  To address this problem
and make the estimation as rapid as the downstream tasks, we borrow methodology
from the popular and successful Vecchia approximation \cite{vecchia1988}, which
is traditionally applied to parametric models (with notable exceptions being
\cite{kidd2020,boyles2021}).

The most standard formulation of Vecchia's approximation is motivated by
elementary observations about probability. In particular, any joint density for
$\by = [Y(\bx_1), \ldots, Y(\bx_n)]$ can be decomposed into a product of
conditional densities as
\begin{equation*} 
  f(y_1, \ldots, y_n) = f(y_1) \prod_{j=2}^n f(y_j \, | \, y_1, \ldots, y_{j-1}).
\end{equation*}
The key observation of Vecchia \cite{vecchia1988} was that in certain settings,
the conditional distribution $y_j \, | \, y_1, \ldots, y_{j-1}$ can be
well-approximated with $y_j \, | \, y_{\sigma(j)_1}, \ldots, y_{\sigma(j)_k}$,
where $\sigma(j) \subset [j-1]$ is a size-$k$ subset of $[j-1]$ that contains the
most highly correlated measurements, and $\sigma(j)_1, \ldots, \sigma(j)_k$
denote its elements. This approximation induces a trivially
$\bO(n)$-cost and embarrassingly parallel approximation to the log-likelihood
given by
\begin{equation} \label{eq:vecchia_small}
  \log f(y_1)
  +
  \sum_{j=2}^n \log f(y_j \, | \, y_1, \ldots, y_{j-1})
  \approx
  \log f(y_1)
  +
  \sum_{j=2}^n \log f(y_j \, | \, y_{\sigma(j)_1}, \ldots, y_{\sigma(j)_k}).
\end{equation}
As a motivating example, a process for which such an approximation is exact is a
Markov process in one dimension with the natural left-to-right ordering, so that
the conditional distribution of $y_j \, | \, y_1, \ldots, y_{j-1}$ is
\emph{exactly} equal to $y_j \, | \, y_{j-1}$. In that way, Vecchia's
approximation can be viewed as an informal Markovian-type approximation. For
parametric Gaussian process models, the accuracy of this approximation is
characterized by the \emph{screening effect} \cite{stein2011}, which can be
heuristically summarized as a specific variety of regularity in the tails of the
Fourier transform of the covariance function \cite{stein2011}. Screening
structure, like Markovian structure, is brittle: separable dependence structure,
measurement noise, and high ambient dimensions can all dramatically reduce the
accuracy of those approximations to conditional distributions
\cite{stein2011,katzfuss2021,schafer2021,geoga2024}. But conversely, regardless
of the setting, Vecchia approximations are optimal with respect to natural
metrics that will be discussed below, and demonstrate consistently strong
performance in practice compared with the wide variety of other fast GP methods,
at least in the setting of spatial data \cite{heaton2019}.

While posing the approximation in terms of small sums as in
\cref{eq:vecchia_small} is both conceptually and computationally convenient,
the approximation can equivalently be viewed as a sparse matrix approximation to
$\bS^{-1}$. In particular, if $\bS^{-1} = \tilde{\bT} \bm{D} \tilde{\bT}\T$,
then $\tilde{\bT}$ may be specified as a unit triangular matrix with
the concrete interpretation that its $j$-th column encodes best linear predictor
weights for $y_j$ given $y_1, \ldots, y_{j-1}$, and $D_{j,j}$ is the inverse
of the conditional variance \cite{pourahmadi1999}. If $\bS^{-1} = \bT
\bT\T$, absorbing $\sqrt{\bm{D}}$ into $\tilde{\bT}$, this interpretation
remains modulo the rescaling by the inverse square-root conditional variance.
Note that with the convention $\sigma(j) \subset [j-1]$ used here, the nonzeros
of column $j$ sit in rows $\sigma(j) \cup \{j\}$, so that $\bT$ is \emph{upper}
triangular; the mirror-image convention gives a lower triangular factor, and the
formulation of \cref{sec:optdesign} accommodates either.
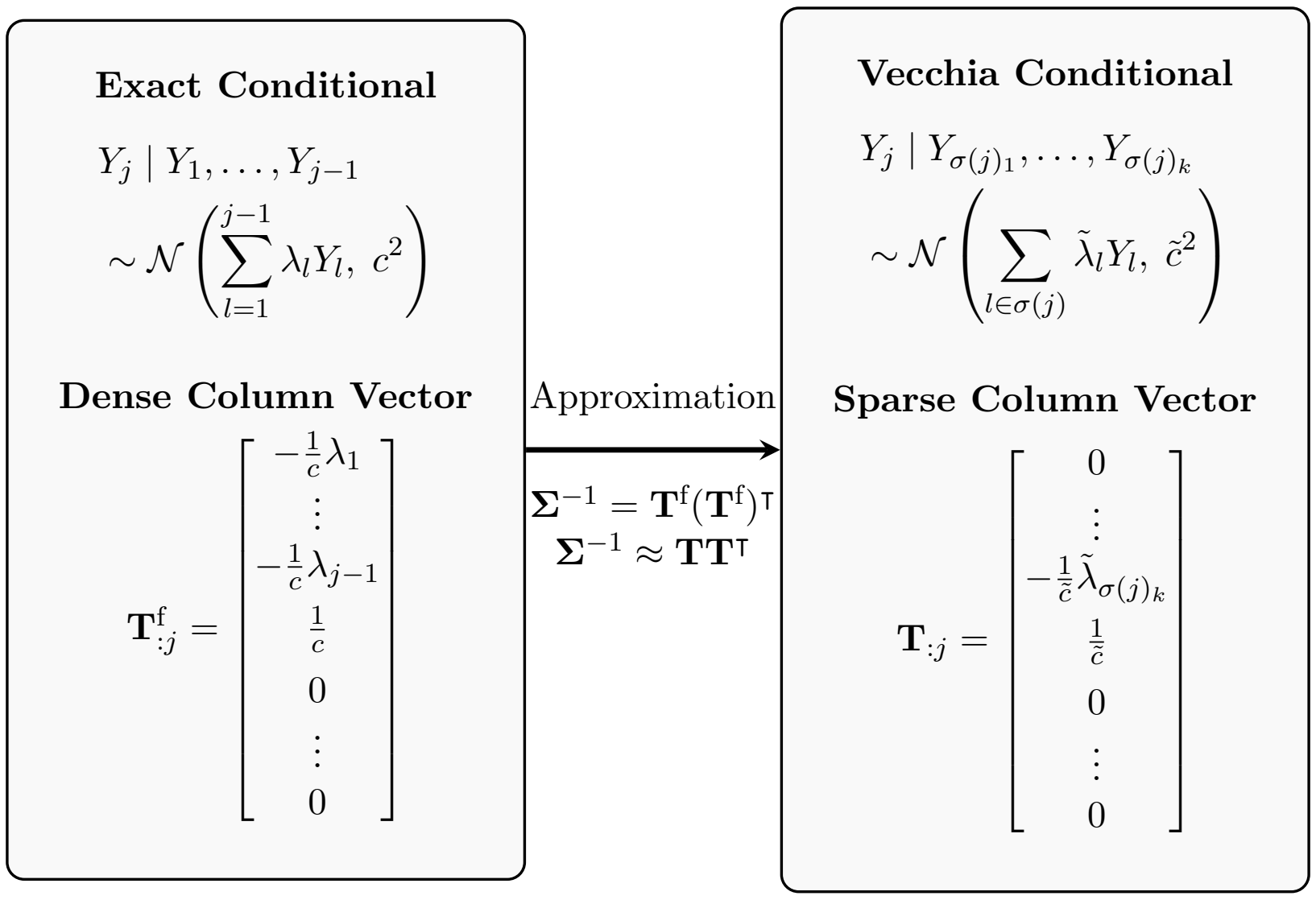
\begin{figure}[!ht]
\centering
\scalebox{0.8}{
\begin{tikzpicture}[
    box/.style={draw, rounded corners=5pt, inner sep=15pt, align=center, fill=gray!5, thick},
    >=stealth
]
    \node[box] (exact_panel) at (0, 0) {
        \textbf{Exact Conditional}\\[4mm]
        $\begin{aligned}
            &Y_j \mid Y_1, \dots, Y_{j-1} \\
            &\sim \mathcal{N}\left(\sum_{l=1}^{j-1} \lambda_l Y_l, \; c^2\right)
        \end{aligned}$ \\[6mm]
        \textbf{Dense Column Vector}\\[2mm]
        $\bT^{\text{f}}_{:j} =
        \begin{bmatrix}
            -\frac{1}{c}\lambda_1 \\
            \vdots \\
            -\frac{1}{c}\lambda_{j-1} \\[1ex]
            \frac{1}{c} \\[1ex]
            0 \\
            \vdots \\
            0
        \end{bmatrix}$
    };
    \node[box] (vecchia_panel) at (8, 0) {
        \textbf{Vecchia Conditional}\\[4mm]
        $\begin{aligned}
            &Y_j \mid Y_{\sigma(j)_1}, \dots, Y_{\sigma(j)_k} \\
            &\sim \mathcal{N}\left(\sum_{l \in \sigma(j)} \tilde{\lambda}_l Y_l,
            \; \tilde{c}^2\right)
        \end{aligned}$ \\[6mm]
        \textbf{Sparse Column Vector}\\[2mm]
        $\bT_{:j} =
        \begin{bmatrix}
            0 \\
            \vdots \\
            -\frac{1}{\tilde{c}}\tilde{\lambda}_{\sigma(j)_k} \\[1ex]
            \frac{1}{\tilde{c}} \\[1ex]
            0 \\
            \vdots \\
            0
        \end{bmatrix}$
    };
    \draw[->, ultra thick] (exact_panel) -- (vecchia_panel)
        node[midway, above, yshift=2mm] {Approximation}
        node[midway, below, yshift=-2mm] {$\bS^{-1} =       \bT^{\text{f}} (\bT^{\text{f}})\T$}
        node[midway, below, yshift=-7mm] {$\bS^{-1} \approx \bT \bT\T$};
\end{tikzpicture}
}
  \caption{A graphical illustration of how the small-sum form of Vecchia
  approximations \cref{eq:vecchia_small} corresponds to a sparse approximation
  to the inverse Cholesky factor of $\bS$ with precisely $|\sigma(j)|$ nonzero
  off-diagonal entries in each column of $\bT$.}
  \label{fig:vecchia_chol}
\end{figure}
\Cref{fig:vecchia_chol} illustrates this property: by specifying the
``prediction sets'' $\{\sigma(j)\}_{j=2}^n$, one has \emph{equivalently}
specified a sparsity pattern to a triangular matrix $\bT$ such that $\bS^{-1}
\approx \bT \bT\T$. If this sparsity pattern is denoted as $\mathcal{S}$,
it has also been shown in \cite{schafer2021} that this method for constructing
$\bT$ is optimal in the Kullback--Leibler sense that
\begin{equation*} 
  \bT_{\text{Vecchia}} =
  \text{argmin}_{\text{nz}(\bT) = \mathcal{S}}
  \;
  D_{\text{KL}}\left(
    \mathcal{N}(\bm{0}, \bS)
    \,
    ||
    \,
    \mathcal{N}(\bm{0}, (\bT \bT\T)^{-1})
  \right),
\end{equation*}
where $\text{nz}(\bT)$ refers to the nonzero indices of $\bT$.

Along with choosing the number of conditioning points $k$, the other two
important design aspects of the Vecchia approximation are (i) the point ordering
and (ii) selecting $k$ points from the full set of prior enumerated points to
condition on. In one dimension, both of these choices are (in most cases)
natural and obvious: points are ordered left-to-right, and the $k$ conditioning
points are chosen as the $k$ most recent time points from the past. In higher
dimensions, there is no such canonical ordering, and so the design questions
become more complicated. Broadly speaking, point enumerations resembling
space-filling designs perform well: the ``maximin'' ordering \cite{guinness2018}
that picks point $\bx_j$ based on maximizing $\min_{1 \leq i < j} ||\bx_j -
\bx_i||$ has been demonstrated to perform well, although particularly for small
$k$ a completely random ordering frequently performs effectively identically
\cite{guinness2018}. And while including far-away points in conditioning sets
has been demonstrated to be advantageous \cite{stein2004}, for both convenience
and intuitive reasons it is overwhelmingly most common to simply pick the $k$
nearest neighbors to point $\bx_j$ when designing $\sigma(j)$
\cite{datta2016,guinness2018,schafer2021}.

These specification choices---an ordering method, which we will choose to be a
random ordering; and a conditioning set design, which we will choose as a simple
$k$-nearest neighbor design---fully imply a nonzero structure to $\bT$. With
this sparsity structure prescribed based on principles from parametric GP
approximations (as opposed to discovered, as is more common in the nonparametric
space) we may view \cref{eq:nll} as a function of the sparse $\bT$, and optimize
it directly with respect to entries of $\bT$. We now move to discussing the
specific formulation of this approach employed here for high-efficiency and
scalable estimation.

\section{Optimization problem design} \label{sec:optdesign}
We define $\bT \in \mathbb{R}^{n \times n}$ as a sparse triangular matrix with $p$ nonzero entries, and aim to determine $\bT$ such that
\begin{equation}
    \label{eq:sparse_inverse}
    \bT\bT\T \approx \bS^{-1}.
\end{equation}
The matrix $\bT$ may be either lower or upper triangular.
From \cref{eq:nll,eq:sparse_inverse}, the negative \emph{log-likelihood} can be expressed in terms of $\bT$ as
\[
\ell(\bT; \; \by) = \frac{1}{2} \left( -2m \log |\bT| + \sum_{s=1}^m \|\bT\T \by_s\|^2 \right).
\]
We denote by $p_j$ the number of nonzeros in the column $j$ of $\bT$, and $p = \sum_{j = 1}^n p_j$ the total number of nonzeros of $\bT$.
To formulate the optimization problem, we introduce two variable vectors: $\bm{x} \in \mathbb{R}^{p}$, which contains the nonzero entries of $\bT$ stored columnwise, and $\bm{w} \in \mathbb{R}^{n}$, which stores the natural logarithms of the diagonal entries of $\bT$.
Introducing these logarithmic variables results in a quadratic objective function that is well defined for all values of the variables, in contrast to a formulation without logarithms, which is only defined on a restricted domain.

The sparsity structure of $\bT$ is described using two arrays, \texttt{rowval} and \texttt{colptr}, of sizes $p$ and $n+1$, respectively.
Together, these arrays encode the locations of the nonzero entries of $\bT$ in compressed sparse column (CSC) format, with the row indices within each column stored in ascending order.
This ordering is consistent with the columnwise storage of $\bm{x}$.
For each column ($j = 1, \ldots, n$), $\text{colptr}[j]$ gives the index of the first nonzero entry in column $j$, while $\text{colptr}[j+1]-1$ gives the index of the last one.
The array \texttt{rowval} stores the corresponding row indices.

From \cref{eq:sparse_inverse} and the definition of $\bm{x}$ and $\bm{w}$, the negative \emph{log-likelihood} can be written as
\[
f(\bm{x}, \bm{w}) =
\sum_{j=1}^n
\left(
- m w_{j}
+
\tfrac{1}{2}
\sum_{s=1}^m
\left(
\sum_{t=\text{colptr}[j]}^{\text{colptr}[j+1]-1}
x_t  y_{s,\text{rowval}[t]}
\right)^2
\right).
\]
To enforce consistency between $\bm{w}$ and the diagonal entries of $\bT$, we introduce the $n$ equality constraints
$\bm{c} : \mathbb{R}^{p} \times \mathbb{R}^{n} \rightarrow \mathbb{R}^n$ defined element-wise by
\[
c_i(\bm{x}, \bm{w}) = e^{w_i} - x_{d_i} = 0,
\quad i = 1, \ldots, n.
\]
Here, $d_i$ denotes the index corresponding to the diagonal element $T_{i,i}$ in $\bm{x}$.
For a lower triangular matrix, $d_i = \text{colptr}[i]$, whereas for an upper triangular matrix,
$d_i = \text{colptr}[i+1]-1$. In addition, bound constraints are imposed on the variables ($w_{1}, \dots, w_{n}$), which are linked to the diagonal entries of $\bT$ through the equality constraints.
This parametrization ensures positivity of the diagonals, allows control over the condition number of $\bT$, and improves statistical stability by preventing excessively small or large diagonal entries.
The resulting optimization problem is
\begin{equation}
  \label{eq:problem}
    \min_{\bm{x} \in \mathbb{R}^{p}, \bm{w} \in \mathbb{R}^{n}} \;  f(\bm{x}, \bm{w})
    \quad \text{subject to}\quad
        \left\{
  \begin{aligned}
    & \bm{c}(\bm{x}, \bm{w}) = 0, \\
      & \bm{l} \le \bm{w} \le \bm{u}\; .
  \end{aligned}
  \right.
\end{equation}
Since $w_i = \log T_{i,i}$, the bounds $\bm{l}$ and $\bm{u}$ are the logarithms of the desired bounds on the diagonal entries and may take either sign; the probabilistic reading of those bounds is given in \cref{sec:opt_reg}.
Since the exponential is increasing and $x_{d_i} = e^{w_i}$ at every feasible point, the constraint $l_i \le w_i \le u_i$ is equivalent to $e^{l_i} \le x_{d_i} \le e^{u_i}$: imposing the bounds on $\bm{w}$ rather than on the diagonal variables $(x_{d_1}, \dots, x_{d_n})$ leaves the feasible set and the solutions of \cref{eq:problem} unchanged.
The choice matters for second-order methods, however.
In an interior-point method (see \cref{sec:opt_kkt}), each bounded variable contributes a diagonal term to the Hessian of the barrier problem, and this term changes at every iteration.
With bounds on $(x_{d_1}, \dots, x_{d_n})$, these terms would be added to the Hessian with respect to $\bm{x}$, whose block-diagonal structure is derived in \cref{sec:opt_derivatives}, and its blocks would then have to be refactorized at every iteration.
With bounds on $\bm{w}$, they only affect a diagonal block of size $n$, and the Hessian with respect to $\bm{x}$ remains constant, which is exploited by the linear solver of \cref{sec:opt_kkt}.
Note that the objective function and the constraints are separable across the columns of $\bT$, with each column-wise subproblem involving only the equality constraint and the bound constraint associated with its diagonal entry.
However, solving the fused formulation retains all constraints simultaneously, which can improve the performance of second-order optimization methods such as interior-point algorithms.
In particular, structure-exploiting linear solvers can take advantage of the block structure of the fused KKT systems (\cref{sec:opt_kkt}) more effectively than generic sparse solvers.

\subsection{Derivatives of the optimization problem} \label{sec:opt_derivatives}

For each column $j$ of $\bT$, let $\bm{x}_{(j)} \in \mathbb{R}^{p_j}$ denote the subvector of $\bm{x}$ that stores the nonzeros of column $j$, i.e., the entries $\text{colptr}[j]$ to $\text{colptr}[j+1]-1$ of $\bm{x}$.
Let $\bm{Y}_j \in \mathbb{R}^{m \times p_j}$ gather the samples at the rows of these nonzeros, i.e., $(\bm{Y}_j)_{s,r} = y_{s,\text{rowval}[\text{colptr}[j]+r-1]}$.
The objective function then reads
\begin{equation}
\label{eq:obj_blocks}
f(\bm{x},\bm{w}) = \sum_{j=1}^n \left( -m w_j + \tfrac{1}{2}\, \bm{x}_{(j)}\T \bm{H}_j \bm{x}_{(j)} \right),
\qquad
\bm{H}_j = \bm{Y}_j\T \bm{Y}_j \in \mathbb{R}^{p_j \times p_j}.
\end{equation}
It is linear in $\bm{w}$ and quadratic in $\bm{x}$, and its derivatives are
\[
\nabla_{\bm{x}_{(j)}} f(\bm{x},\bm{w}) = \bm{H}_j \bm{x}_{(j)}, \qquad
\nabla_{\bm{w}} f(\bm{x},\bm{w}) = -m \bm{e}, \qquad
\nabla^2 f(\bm{x},\bm{w}) =
\begin{bmatrix} \bm{H} & 0 \\ 0 & 0 \end{bmatrix},
\]
where $\bm{e} \in \mathbb{R}^n$ is the vector of ones and $\bm{H} = \diag(\bm{H}_1, \ldots, \bm{H}_n) \in \mathbb{R}^{p \times p}$ is block diagonal.
The blocks $\bm{H}_j$ are symmetric positive semidefinite and do not depend on $(\bm{x}, \bm{w})$.
They are therefore computed once, with
\[
(\bm{H}_j)_{r,c} = \sum_{s=1}^m y_{s,\text{rowval}[\text{colptr}[j]+r-1]} \; y_{s,\text{rowval}[\text{colptr}[j]+c-1]}, \qquad 1 \le r, c \le p_j,
\]
after which $f$ and $\nabla f$ are evaluated with \cref{eq:obj_blocks} in $\bO(\sum_j p_j^2)$ operations, independently of the number of replicates $m$.
Only the lower triangular part of each block is stored, columnwise, which amounts to $\sum_j p_j (p_j+1)/2$ coefficients.

The constraints are linear in $\bm{x}$ and separable in $\bm{w}$.
Their Jacobian is $[\bm{J} \; \bm{D}]$, where $\bm{J} \in \mathbb{R}^{n \times p}$ is the Jacobian with respect to $\bm{x}$, whose $i$-th row is $-\bm{e}_{d_i}\T$ with $\bm{e}_{d_i}$ the $d_i$-th column of the identity matrix $\bm{I}_p$, and $\bm{D} = \diag(e^{w_1}, \ldots, e^{w_n})$ is the Jacobian with respect to $\bm{w}$.
The only nonzero second-order derivative of $c_i$ is $\partial^2 c_i / \partial w_i^2 = e^{w_i}$, so that, for any $\bmsym{\lambda} \in \mathbb{R}^n$,
\[
\sum_{i=1}^n \lambda_i \nabla^2 c_i(\bm{x},\bm{w}) =
\begin{bmatrix} 0 & 0 \\ 0 & \diag(\lambda_1 e^{w_1}, \ldots, \lambda_n e^{w_n}) \end{bmatrix}.
\]

\subsection{KKT systems and specialized linear solver} \label{sec:opt_kkt}

Let $\bmsym{\lambda} \in \mathbb{R}^n$ denote the Lagrange multipliers associated with the equality constraints $\bm{c}(\bm{x},\bm{w}) = 0$, and let $(\bm{z}_l, \bm{z}_u) \in \mathbb{R}^n \times \mathbb{R}^n$ denote the multipliers associated with the lower and upper bound constraints on $\bm{w}$, respectively.
Using the primal and dual variables, the Lagrangian of \cref{eq:problem} is defined as
\begin{align*}
\mathcal{L}(\bm{x},\bm{w},\bmsym{\lambda},\bm{z}_l,\bm{z}_u)
= f(\bm{x},\bm{w}) + \bmsym{\lambda}\T \bm{c}(\bm{x},\bm{w}) - \bm{z}_l\T (\bm{w} - \bm{l}) - \bm{z}_u\T (\bm{u} - \bm{w}).
\end{align*}

The KKT conditions of \cref{eq:problem} are:
\begin{equation}
    \label{eq:kkt_conditions}
\begin{cases}
    \nabla_{\bm{x}} f(\bm{x},\bm{w}) + \sum_{i=1}^n \lambda_i \nabla_{\bm{x}} c_i(\bm{x},\bm{w}) = 0 \; , \\
    \nabla_{\bm{w}} f(\bm{x},\bm{w}) + \sum_{i=1}^n \lambda_i \nabla_{\bm{w}} c_i(\bm{x},\bm{w}) - \bm{z}_l + \bm{z}_u = 0 \; , \\
    \bm{c}(\bm{x},\bm{w}) = 0 \; , \\
    0 \leq \bm{z}_l \perp \bm{w} - \bm{l} \geq 0 \; , \\
    0 \leq \bm{z}_u \perp \bm{u} - \bm{w} \geq 0 \; .
\end{cases}
\end{equation}
The notation $\bm{a} \perp \bm{b}$ is a shorthand for the complementarity
condition $a_i b_i = 0$ (for all $i=1,\ldots, n$).
The interior-point method aims at finding a stationary point
satisfying the KKT conditions \cref{eq:kkt_conditions}.
The complementarity constraints render the KKT conditions non-smooth, complicating the solution of
the whole system \cref{eq:kkt_conditions}.
Interior-point methods use a homotopy continuation method to solve a simplified
version of \cref{eq:kkt_conditions}, parameterized by a barrier
parameter $\mu > 0$.
For $(\bm{w} - \bm{l}, \bm{u} - \bm{w}, \bm{z}_l, \bm{z}_u) > 0$, we solve the system
\begin{equation}
  \label{eq:kkt_ipm}
  \bm{F}_\mu(\bm{x}, \bm{w}, \bmsym{\lambda}, \bm{z}_l, \bm{z}_u) =
  \begin{bmatrix}
       \nabla_{\bm{x}} f(\bm{x},\bm{w}) + \nabla_{\bm{x}} \bm{c}(\bm{x},\bm{w})\T \bmsym{\lambda}  \\
       \nabla_{\bm{w}} f(\bm{x},\bm{w}) + \nabla_{\bm{w}} \bm{c}(\bm{x},\bm{w})\T \bmsym{\lambda} - \bm{z}_l + \bm{z}_u \\
       \bm{c}(\bm{x},\bm{w})  \\
       \bm{Z}_l (\bm{W} - \bm{L}) \bm{e} - \mu \bm{e} \\
       \bm{Z}_u (\bm{U} - \bm{W}) \bm{e} - \mu \bm{e}
  \end{bmatrix} = 0
   \; .
\end{equation}
In \cref{eq:kkt_ipm}, $\bm{W}$, $\bm{L}$, $\bm{U}$, $\bm{Z}_l$ and $\bm{Z}_u$ denote the diagonal matrices whose diagonals are $\bm{w}$, $\bm{l}$, $\bm{u}$, $\bm{z}_l$ and $\bm{z}_u$, respectively.
As we drive the barrier parameter $\mu$ to 0, the solution of the
system $\bm{F}_\mu(\bm{x}, \bm{w}, \bmsym{\lambda}, \bm{z}_l, \bm{z}_u) = 0$ tends to the solution of the
KKT conditions \cref{eq:kkt_conditions}.
We note that for a fixed barrier parameter $\mu$, the function $\bm{F}_\mu(\cdot)$ is smooth.
Hence, the system \cref{eq:kkt_ipm} can be solved iteratively using a standard Newton method.
At each iteration, the Newton step
\begin{equation}
\label{eq:newton_step}
\nabla_{\bm{t}} \bm{F}_\mu(\bm{t}) \Delta \bm{t} = - \bm{F}_\mu(\bm{t}), \quad \bm{t} = (\bm{x}, \bm{w}, \bmsym{\lambda}, \bm{z}_l, \bm{z}_u),
\end{equation}
provides a search direction $\Delta \bm{t}$ that is used to update the current iterate toward the solution of the perturbed KKT conditions.
Once the iterates are sufficiently close to the central path,
the optimization solver decreases the barrier parameter $\mu$ to find a solution closer to
the original KKT conditions.
In interior-point methods, the bulk of the workload is the computation of the Newton
step \cref{eq:newton_step}, which involves assembling the Jacobian
$\nabla_{\bm{t}} \bm{F}_\mu(\bm{t})$ of the KKT conditions and solving the resulting linear system to compute
the descent direction $\Delta \bm{t}$.
By writing out all the blocks, the system in \cref{eq:newton_step} has the following structure:
\begin{equation}
  \label{eq:kkt:unreduced}
  \begin{bmatrix}
    \bm{H} & 0 & \bm{J}\T & 0 & 0 \\
    0 & \bm{\Lambda} \bm{D} & \bm{D}\T  & -\bm{I} & \bm{I} \\
    \bm{J} & \bm{D} & 0 & 0 & 0 \\
    0 & \bm{Z}_l & 0 & \bm{W} - \bm{L} & 0 \\
    0 & -\bm{Z}_u & 0 & 0 & \bm{U} - \bm{W}
  \end{bmatrix}
  \begin{bmatrix}
    \Delta \bm{x} \\
    \Delta \bm{w} \\
    \Delta \bmsym{\lambda} \\
    \Delta \bm{z}_l \\
    \Delta \bm{z}_u
  \end{bmatrix}
  = - \begin{bmatrix}
  \nabla_{\bm{x}} f(\bm{x},\bm{w}) + \bm{J}\T \bmsym{\lambda} \\
  -m\bm{e} + \bm{D}\T \bmsym{\lambda} - \bm{z}_l + \bm{z}_u \\
  \bm{c}(\bm{x},\bm{w})  \\
  \bm{Z}_l (\bm{W} - \bm{L}) \bm{e} - \mu \bm{e} \\
  \bm{Z}_u (\bm{U} - \bm{W}) \bm{e} - \mu \bm{e}
  \end{bmatrix},
\end{equation}
where $\bm{\Lambda} = \diag(\lambda_1, \dots, \lambda_n) \in \mathbb{R}^{n \times n}$, and $\bm{H}$, $\bm{J}$ and $\bm{D}$ are defined in \cref{sec:opt_derivatives}.
The Jacobian $\bm{J}$ and the Hessian $\bm{H}$ are constant across iterations, whereas $\bm{\Lambda}$, $\bm{D}$, $\bm{Z}_l$, $\bm{Z}_u$, $\bm{W}-\bm{L}$ and $\bm{U}-\bm{W}$ depend on the current iterate.
If we eliminate $\Delta \bm{z}_l$ and $\Delta \bm{z}_u$ with the following equations:
\begin{align*}
\Delta \bm{z}_l &= -\bm{z}_l - (\bm{W} - \bm{L})^{-1} \bm{Z}_l \Delta \bm{w} + \mu (\bm{W} - \bm{L})^{-1} \bm{e}, \\
\Delta \bm{z}_u &= -\bm{z}_u + (\bm{U} - \bm{W})^{-1} \bm{Z}_u \Delta \bm{w} + \mu (\bm{U} - \bm{W})^{-1} \bm{e},
\end{align*}
we obtain:
\begin{equation}
  \label{eq:kkt:reduced}
  \begin{bmatrix}
    \bm{H} & 0 & \bm{J}\T \\
    0 & \bm{Q} & \bm{D}\T \\
    \bm{J} & \bm{D} & 0
  \end{bmatrix}
  \begin{bmatrix}
    \Delta \bm{x} \\
    \Delta \bm{w} \\
    \Delta \bmsym{\lambda}
  \end{bmatrix}
  = - \begin{bmatrix}
  \nabla_{\bm{x}} f(\bm{x},\bm{w}) + \bm{J}\T \bmsym{\lambda} \\
  \bm{b}_w \\
  \bm{c}(\bm{x},\bm{w})
  \end{bmatrix},
\end{equation}
where
\begin{align*}
  \bm{Q} &= \bm{\Lambda} \bm{D} + (\bm{W} -\bm{L})^{-1} \bm{Z}_l + (\bm{U} - \bm{W})^{-1} \bm{Z}_u, \\
  \bm{b}_w &= -m\bm{e} + \bm{D}\T \bmsym{\lambda} - \mu (\bm{W} - \bm{L})^{-1} \bm{e} + \mu (\bm{U} - \bm{W})^{-1} \bm{e}.
\end{align*}
To further reduce the system, we eliminate $\Delta \bm{x}$, which requires $\bm{H}$ to be invertible.
Whether $\bm{H}$ is invertible is governed by the number of replicates: since $\bm{H}_j = \bm{Y}_j\T \bm{Y}_j$ with $\bm{Y}_j \in \mathbb{R}^{m \times p_j}$, we have $\operatorname{rank} \bm{H}_j \le m$, and $\bm{H}_j$ is singular whenever $m < p_j$, where $p_j = k+1$ for all but the first $k$ columns.
The estimation problem therefore requires at least $k+1$ replicates per column to be well posed without regularization, and the relevant sample-size ratio is $m/(k+1)$ per column rather than $m$ against the $n(k+1)$ unknowns of the full problem---a consequence of the column separability noted above.
Since $\bm{H}$ is in general not guaranteed to be invertible, a standard approach is to add $\gamma \bm{I}_p$ to $\bm{H}$, which ensures invertibility for $\gamma > 0$.
This is equivalent to adding the ridge penalty $\tfrac{\gamma}{2} \|\bm{x}\|^2$ to the objective function of \cref{eq:problem}, whose statistical interpretation is discussed in \cref{sec:opt_reg}.

When $\bm{H}$ is invertible, the first block row of \cref{eq:kkt:reduced} gives
\begin{equation}
\label{eq:dx}
\Delta \bm{x} = - \bm{H}^{-1} \left( \nabla_{\bm{x}} f(\bm{x}, \bm{w}) + \bm{J}\T(\bmsym{\lambda} + \Delta \bmsym{\lambda}) \right).
\end{equation}
Eliminating $\Delta \bm{x}$ from \cref{eq:kkt:reduced} with \cref{eq:dx} amounts to forming the Schur complement of $\bm{H}$ in the matrix of \cref{eq:kkt:reduced},
\[
\bm{S}
=
\begin{bmatrix}
\bm{Q} & \bm{D}\T \\
\bm{D} & 0
\end{bmatrix}
-
\begin{bmatrix}
0 \\
\bm{J}
\end{bmatrix}
\bm{H}^{-1}
\begin{bmatrix}
0 & \bm{J}\T
\end{bmatrix}
=
\begin{bmatrix}
\bm{Q} & \!\! \phantom{-}\bm{D}\T \\
\bm{D} & \!\! -\bm{M}
\end{bmatrix},
\]
where $\bm{M} = \bm{J} \bm{H}^{-1} \bm{J}\T$ is diagonal with diagonal coefficients $M_{i,i} = H^{-1}_{d_i,d_i}$, since each row of $\bm{J}$ selects a single entry of $\bm{x}$.
We obtain the condensed KKT system
\begin{equation*}
  \bm{S}
  \begin{bmatrix}
    \Delta \bm{w} \\
    \Delta \bmsym{\lambda}
  \end{bmatrix}
  = - \begin{bmatrix}
  \bm{b}_w \\
  \bm{b}_\lambda
  \end{bmatrix},
\end{equation*}
where $\bm{b}_w$ is unchanged, since the second block row of \cref{eq:kkt:reduced} does not involve $\Delta \bm{x}$, and
\[
  \bm{b}_\lambda = \bm{c}(\bm{x},\bm{w}) - \bm{J} \bm{H}^{-1} \nabla_{\bm{x}} f(\bm{x}, \bm{w}) - \bm{M}\bmsym{\lambda}.
\]
Since $\bm{Q}$, $\bm{D}$ and $\bm{M}$ are diagonal, this system decouples into $n$ independent $2 \times 2$ systems
\[
  \begin{bmatrix}
    Q_{i,i} & \phantom{-}D_{i,i} \\
    D_{i,i} & -M_{i,i}
  \end{bmatrix}
  \begin{bmatrix}
    \Delta w_i \\
    \Delta \lambda_i
  \end{bmatrix}
  = - \begin{bmatrix}
  b_{w,i} \\
  b_{\lambda,i}
  \end{bmatrix},
  \quad i = 1, \ldots, n,
\]
whose determinants are $-\eta_i$, with $\eta_i = Q_{i,i} M_{i,i} + D_{i,i}^2$.
Their solutions are
\begin{align*}
\Delta w_i &= -\frac{M_{i,i}\, b_{w,i} + D_{i,i}\, b_{\lambda,i}}{\eta_i}, &
\Delta \lambda_i &= \frac{Q_{i,i}\, b_{\lambda,i} - D_{i,i}\, b_{w,i}}{\eta_i},
\end{align*}
which cost $\bO(n)$ operations in total.
We note that $\bm{Q}$ does not need to be invertible: $Q_{i,i}$ may vanish or be negative, since the multiplier $\lambda_i$ has no sign constraint.
The formulas above only require $\eta_i \neq 0$, which holds as soon as the KKT matrix has the inertia expected by the interior-point method, as we show next.

The block elimination also provides the inertia of the KKT matrix in \cref{eq:kkt:reduced}, which interior-point methods monitor to detect directions of negative curvature.
By the Haynsworth inertia additivity formula, if $\bm{H} \succ 0$, the inertia of the matrix in \cref{eq:kkt:reduced} is the sum of the inertia of $\bm{H}$, that is $p$ positive eigenvalues, and of the inertia of $\bm{S}$.
After a symmetric permutation, $\bm{S}$ is block diagonal with the $n$ blocks of size $2 \times 2$ defined above.
A symmetric $2 \times 2$ matrix has one positive and one negative eigenvalue if and only if its determinant is negative.
Hence the KKT matrix has the inertia $(p+n, 0, n)$ expected by the interior-point method if and only if $\eta_i > 0$ for $i = 1, \ldots, n$, which is verified in $\bO(n)$ operations, without any sparse symmetric indefinite factorization.
In particular, the $2 \times 2$ systems are nonsingular whenever the inertia is correct.
If the Cholesky factorization of a block $\bm{H}_j$ fails, or if $\eta_i \le 0$ for some $i$, the inertia is reported as incorrect and the interior-point method regularizes the system, as discussed below.

Note that it is not necessary to explicitly compute $\bm{H}^{-1}$.
In practice, the blocks $\bm{H}_j$ are expected to be dense and are factorized with a Cholesky factorization $\bm{H}_j = \bm{C}_j \bm{C}_j\T$, with $\bm{C}_j$ lower triangular; a product $\bm{q} = \bm{H}^{-1} \bm{r}$ is then computed with two triangular solves per block.
Since the $n$ blocks are small and independent, these factorizations and solves are batched: each block is kept in the packed storage of the Hessian (its lower triangular part, stored columnwise), and a single thread factorizes it and performs its triangular solves, on CPUs and GPUs alike.
Right after the factorization of $\bm{H}_j$, we also compute the column $\bm{H}_j^{-1} \bm{e}_{d_j}$ associated with the diagonal entry of column $j$ of $\bT$.
These columns provide the diagonal coefficients of $\bm{M}$, and since $\bm{J}\T \Delta \bmsym{\lambda} = -\sum_{i=1}^n \Delta \lambda_i \bm{e}_{d_i}$, the term $\bm{H}^{-1} \bm{J}\T \Delta \bmsym{\lambda}$ in \cref{eq:dx} is a linear combination of them.
Each Newton step therefore requires a single solve with the blocks $\bm{H}_j$, for the right-hand side $\nabla_{\bm{x}} f(\bm{x},\bm{w}) + \bm{J}\T \bmsym{\lambda}$.
The particular block structure of the KKT system can be exploited to circumvent the use of sparse linear solvers.

We implemented the formulation \cref{eq:problem} in the open-source Julia~\cite{bezanson2017} package NonparametricVecchia.jl\footnote{\texttt{https://github.com/exanauts/NonparametricVecchia.jl}}, and we solve it with the interior-point solver MadNLP~\cite{shin2021graph,shin2024accelerating}.
The block elimination described in this section is available in the package as \texttt{VecchiaKKTSystem}, a custom KKT system for MadNLP that replaces the sparse factorization of \cref{eq:kkt:reduced} by the Cholesky factorizations of the blocks $\bm{H}_j$ and the $n$ independent $2 \times 2$ systems above.
It runs on CPUs, as well as on NVIDIA and AMD GPUs through the packages CUDA.jl~\cite{besard2018juliagpu} and AMDGPU.jl\footnote{\texttt{https://github.com/JuliaGPU/AMDGPU.jl}}: the factorizations and the solves with the blocks $\bm{H}_j$ are performed by vendor-agnostic kernels written with KernelAbstractions.jl~\cite{churavy_kernelabstractions}, and the objective and its derivatives are also evaluated on the GPU.

The derivation above ignores the modifications that interior-point solvers apply to the Newton system.
When the inertia is not the expected one, the primal block is regularized by $\delta_p \bm{I}$ and the zero block by $-\delta_d \bm{I}$; the problem may also be scaled.
None of these modifications affects the structure exploited above: $\delta_p$ is added to the diagonal of the blocks $\bm{H}_j$ and to $\bm{Q}$, $-\delta_d$ is added to the diagonal of $-\bm{M}$, and the scaling rescales these blocks without changing their structure.
In our implementation, a block $\bm{H}_j$ is refactorized only when its entries or its regularization have changed since its last factorization.

It is worth recording the resulting complexity.
With $k$ conditioning points per column we have $p_j = k+1$, so assembling the $n$ blocks $\bm{H}_j$ costs $\bO(n m k^2)$ and factorizing them costs $\bO(n k^3)$; both are paid once, since $\bm{H}$ does not depend on the iterate (up to a refactorization whenever the regularization $\delta_p$ changes).
Each interior-point iteration then costs $\bO(n k^2)$ for the triangular solves with the $\bm{H}_j$ plus $\bO(n)$ for the diagonal operations above.
The total is therefore $\bO(n m k^2 + n k^3 + N_{\text{it}} n k^2)$, linear in $n$ at fixed $k$.
For the regimes considered in \cref{sec:demo}, where $m \gg k$, the one-time assembly of $\bm{H}$ is the dominant term.

\subsection{Convexity of the reformulation} \label{sec:opt_convexity}

We now return to the convexity of \cref{eq:problem}, since the two formulations available to us differ in this respect.
Written directly in terms of the nonzeros of $\bT$, the estimation problem is convex: $-2m \log |\bT| + \sum_{s} \|\bT\T \by_s\|^2$ is a sum of a convex quadratic and the convex barrier $-2m \sum_j \log T_{j,j}$ on the open set $\{T_{j,j} > 0\}$, and the box constraints on the diagonal are linear.
Problem \cref{eq:problem} is a smooth reparameterization of that convex problem through the change of variables $T_{i,i} = e^{w_i}$.
The reparameterization trades convexity of the feasible set---the equality constraints $e^{w_i} - x_{d_i} = 0$ are nonlinear---for an objective that is quadratic and globally defined, and for bound constraints that no longer interact with the logarithmic singularity.
The change of variables is a diffeomorphism between the two feasible sets, and the Jacobian $[\bm{J} \; \bm{D}]$ of the equality constraints has full row rank $n$ because $\bm{D}$ is nonsingular, so LICQ holds at every feasible point and the KKT points of the two problems are in bijection.
What is not inherited automatically is positive semidefiniteness of the Hessian of the Lagrangian, since the constraints now contribute curvature $\lambda_i e^{w_i}$.
That curvature is nevertheless pinned down by stationarity: the stationarity condition in $w_i$ of \cref{eq:kkt_conditions} reads $-m + \lambda_i e^{w_i} - z_{l,i} + z_{u,i} = 0$, so at any point satisfying the stationarity conditions---in particular at every KKT point and at every point of the barrier central path---
\begin{equation} \label{eq:curvature}
  \nabla^2 \mathcal{L} =
  \begin{bmatrix} \bm{H} & 0 \\ 0 & \diag(m + z_{l,1} - z_{u,1}, \ldots, m + z_{l,n} - z_{u,n}) \end{bmatrix}.
\end{equation}
Since $\bm{H} \succeq 0$, the Lagrangian Hessian is positive semidefinite unless some upper bound is active with multiplier $z_{u,i} > m$; complementarity forbids $z_{l,i}$ and $z_{u,i}$ from being simultaneously positive, so no other configuration can produce negative curvature.
An active upper bound with $z_{u,i} > m$ means the data call for a larger inverse conditional standard deviation than the box of \cref{sec:opt_reg} permits, which is a signal to widen the box rather than a difficulty intrinsic to the formulation.
Outside that regime the problem behaves as the convex one it reparameterizes, and the inertia correction described in \cref{sec:opt_kkt} covers the remainder; we have not observed it to be a limiting factor.

\subsection{Regularization} \label{sec:opt_reg}

Due to the tight connection between the inverse symmetric factor $\bT$ and
the implied probabilistic model for the data, certain forms of regularization
are particularly natural. In this section, we discuss two such choices that also
can be valuable in stabilizing the above optimization procedure. Both preserve
convexity of the underlying problem in $\bT$.

First, we note that a box constraint on the diagonal $\{T_{j,j}\}_{j=1}^n$
serves the important numerical purpose of controlling the condition number of
$\bT$, and thus the sparsely approximated $\bS^{-1}$. Probabilistically, this
has a similar concrete interpretation as constraining the implied inverse
square-root conditional variance. Consider, for example, a process with
spatially fixed marginal variance $\varsigma^2$ that is presumed to have
uncorrelated measurement noise with variance $\tau^2$. Naturally, then, this
constrains the conditional variances for any prediction problem to be in the
interval $[\tau^2, \varsigma^2]$, and so such a condition can naturally be enforced
by constraining that $T_{j,j} \in [\varsigma^{-1}, \tau^{-1}]$, or equivalently
$w_j \in [-\log \varsigma, -\log \tau]$ in the variables of \cref{eq:problem}.

A related interpretation can be applied to the off-diagonal entries. Recalling
(see \cref{fig:vecchia_chol}) that columns of $\bT$ can be interpreted
as rescaled negative Kriging weights, the ridge-type penalty term
$\tfrac{\gamma}{2} ||\bT||_{F}^2 = \tfrac{\gamma}{2} \|\bm{x}\|^2$ of \cref{sec:opt_kkt} effectively works to reduce the norm
of the prediction weights and raise the conditional variance. The penalty also
acts on the diagonal entries $T_{j,j}$, whose scale is however already
controlled by the box constraint above; including them makes each block
$\bm{H}_j + \gamma \bm{I}$ positive definite even when $m < k+1$, which is
the purpose of this term in \cref{sec:opt_kkt}. Because this term
is a convex quadratic in $\bm{x}$, it leaves the
structure of \cref{eq:problem} untouched: it adds the constant diagonal shift
$\gamma \bm{I}$ to each block $\bm{H}_j$ and so is absorbed at no cost by the
solver described above. This property of
prediction weights is closely connected to the roughness of a GP. This intuition
can be seen in the following way: consider predicting $Y(\bx_0)$ given $\by$ for
a mean-zero process. The Kriging weights will be given by $\bS_{\by}^{-1}
\bmsym{\kappa}$, where $\bmsym{\kappa}$ is the vector of cross-covariances with
entries $\kappa_j = K(\bx_j, \bx_0)$.
And so if $\bS$ is poorly conditioned (for example when the covariance function
$K$ is globally smooth), $|| \bS^{-1} \bmsym{\kappa}||$ may be very large if
$\bmsym{\kappa}$
has content from the near-degenerate singular value space of $\bS$. If $\bS$ is
well-conditioned, on the other hand, $|| \bS^{-1} \bmsym{\kappa}||$ will naturally be
smaller. For more background connecting smoothness properties of $K$ with
algebraic properties of prediction weights, we refer readers to \cite{stein1999}.

\section{Numerical demonstrations} \label{sec:demo}

\subsection{Simulation study} \label{sec:simstudy}

\begin{figure}[!ht]
  \centering
\begingroup
  \makeatletter
  \providecommand\color[2][]{%
    \GenericError{(gnuplot) \space\space\space\@spaces}{%
      Package color not loaded in conjunction with
      terminal option `colourtext'%
    }{See the gnuplot documentation for explanation.%
    }{Either use 'blacktext' in gnuplot or load the package
      color.sty in LaTeX.}%
    \renewcommand\color[2][]{}%
  }%
  \providecommand\includegraphics[2][]{%
    \GenericError{(gnuplot) \space\space\space\@spaces}{%
      Package graphicx or graphics not loaded%
    }{See the gnuplot documentation for explanation.%
    }{The gnuplot epslatex terminal needs graphicx.sty or graphics.sty.}%
    \renewcommand\includegraphics[2][]{}%
  }%
  \providecommand\rotatebox[2]{#2}%
  \@ifundefined{ifGPcolor}{%
    \newif\ifGPcolor
    \GPcolortrue
  }{}%
  \@ifundefined{ifGPblacktext}{%
    \newif\ifGPblacktext
    \GPblacktexttrue
  }{}%
  \let\gplgaddtomacro\g@addto@macro
  \gdef\gplbacktext{}%
  \gdef\gplfronttext{}%
  \makeatother
  \ifGPblacktext
    \def\colorrgb#1{}%
    \def\colorgray#1{}%
  \else
    \ifGPcolor
      \def\colorrgb#1{\color[rgb]{#1}}%
      \def\colorgray#1{\color[gray]{#1}}%
      \expandafter\def\csname LTw\endcsname{\color{white}}%
      \expandafter\def\csname LTb\endcsname{\color{black}}%
      \expandafter\def\csname LTa\endcsname{\color{black}}%
      \expandafter\def\csname LT0\endcsname{\color[rgb]{1,0,0}}%
      \expandafter\def\csname LT1\endcsname{\color[rgb]{0,1,0}}%
      \expandafter\def\csname LT2\endcsname{\color[rgb]{0,0,1}}%
      \expandafter\def\csname LT3\endcsname{\color[rgb]{1,0,1}}%
      \expandafter\def\csname LT4\endcsname{\color[rgb]{0,1,1}}%
      \expandafter\def\csname LT5\endcsname{\color[rgb]{1,1,0}}%
      \expandafter\def\csname LT6\endcsname{\color[rgb]{0,0,0}}%
      \expandafter\def\csname LT7\endcsname{\color[rgb]{1,0.3,0}}%
      \expandafter\def\csname LT8\endcsname{\color[rgb]{0.5,0.5,0.5}}%
    \else
      \def\colorrgb#1{\color{black}}%
      \def\colorgray#1{\color[gray]{#1}}%
      \expandafter\def\csname LTw\endcsname{\color{white}}%
      \expandafter\def\csname LTb\endcsname{\color{black}}%
      \expandafter\def\csname LTa\endcsname{\color{black}}%
      \expandafter\def\csname LT0\endcsname{\color{black}}%
      \expandafter\def\csname LT1\endcsname{\color{black}}%
      \expandafter\def\csname LT2\endcsname{\color{black}}%
      \expandafter\def\csname LT3\endcsname{\color{black}}%
      \expandafter\def\csname LT4\endcsname{\color{black}}%
      \expandafter\def\csname LT5\endcsname{\color{black}}%
      \expandafter\def\csname LT6\endcsname{\color{black}}%
      \expandafter\def\csname LT7\endcsname{\color{black}}%
      \expandafter\def\csname LT8\endcsname{\color{black}}%
    \fi
  \fi
    \setlength{\unitlength}{0.0500bp}%
    \ifx\gptboxheight\undefined%
      \newlength{\gptboxheight}%
      \newlength{\gptboxwidth}%
      \newsavebox{\gptboxtext}%
    \fi%
    \setlength{\fboxrule}{0.5pt}%
    \setlength{\fboxsep}{1pt}%
    \definecolor{tbcol}{rgb}{1,1,1}%
\begin{picture}(7360.00,3960.00)%
    \gplgaddtomacro\gplbacktext{%
      \csname LTb\endcsname
      \put(266,2169){\makebox(0,0)[r]{\strut{}\scriptsize 0}}%
      \csname LTb\endcsname
      \put(266,2444){\makebox(0,0)[r]{\strut{}\scriptsize 0.2}}%
      \csname LTb\endcsname
      \put(266,2719){\makebox(0,0)[r]{\strut{}\scriptsize 0.4}}%
      \csname LTb\endcsname
      \put(266,2993){\makebox(0,0)[r]{\strut{}\scriptsize 0.6}}%
      \csname LTb\endcsname
      \put(266,3268){\makebox(0,0)[r]{\strut{}\scriptsize 0.8}}%
      \csname LTb\endcsname
      \put(266,3543){\makebox(0,0)[r]{\strut{}\scriptsize 1}}%
    }%
    \gplgaddtomacro\gplfronttext{%
      \csname LTb\endcsname
      \put(1382,3725){\makebox(0,0){\strut{}\footnotesize real data sample}}%
    }%
    \gplgaddtomacro\gplbacktext{%
    }%
    \gplgaddtomacro\gplfronttext{%
      \csname LTb\endcsname
      \put(3669,3725){\makebox(0,0){\strut{}\footnotesize real data sample}}%
    }%
    \gplgaddtomacro\gplbacktext{%
    }%
    \gplgaddtomacro\gplfronttext{%
      \csname LTb\endcsname
      \put(7225,2167){\makebox(0,0)[l]{\strut{}\tiny -3}}%
      \csname LTb\endcsname
      \put(7225,2396){\makebox(0,0)[l]{\strut{}\tiny -2}}%
      \csname LTb\endcsname
      \put(7225,2626){\makebox(0,0)[l]{\strut{}\tiny -1}}%
      \csname LTb\endcsname
      \put(7225,2856){\makebox(0,0)[l]{\strut{}\tiny 0}}%
      \csname LTb\endcsname
      \put(7225,3086){\makebox(0,0)[l]{\strut{}\tiny 1}}%
      \csname LTb\endcsname
      \put(7225,3316){\makebox(0,0)[l]{\strut{}\tiny 2}}%
      \csname LTb\endcsname
      \put(7225,3545){\makebox(0,0)[l]{\strut{}\tiny 3}}%
      \csname LTb\endcsname
      \put(5957,3725){\makebox(0,0){\strut{}\footnotesize real data sample}}%
    }%
    \gplgaddtomacro\gplbacktext{%
      \csname LTb\endcsname
      \put(266,396){\makebox(0,0)[r]{\strut{}\scriptsize 0}}%
      \csname LTb\endcsname
      \put(266,671){\makebox(0,0)[r]{\strut{}\scriptsize 0.2}}%
      \csname LTb\endcsname
      \put(266,946){\makebox(0,0)[r]{\strut{}\scriptsize 0.4}}%
      \csname LTb\endcsname
      \put(266,1220){\makebox(0,0)[r]{\strut{}\scriptsize 0.6}}%
      \csname LTb\endcsname
      \put(266,1495){\makebox(0,0)[r]{\strut{}\scriptsize 0.8}}%
      \csname LTb\endcsname
      \put(266,1770){\makebox(0,0)[r]{\strut{}\scriptsize 1}}%
      \csname LTb\endcsname
      \put(371,250){\makebox(0,0){\strut{}\scriptsize 0}}%
      \csname LTb\endcsname
      \put(775,250){\makebox(0,0){\strut{}\scriptsize 0.2}}%
      \csname LTb\endcsname
      \put(1180,250){\makebox(0,0){\strut{}\scriptsize 0.4}}%
      \csname LTb\endcsname
      \put(1584,250){\makebox(0,0){\strut{}\scriptsize 0.6}}%
      \csname LTb\endcsname
      \put(1989,250){\makebox(0,0){\strut{}\scriptsize 0.8}}%
      \csname LTb\endcsname
      \put(2393,250){\makebox(0,0){\strut{}\scriptsize 1}}%
    }%
    \gplgaddtomacro\gplfronttext{%
      \csname LTb\endcsname
      \put(1382,1952){\makebox(0,0){\strut{}\footnotesize fitted model simulation}}%
    }%
    \gplgaddtomacro\gplbacktext{%
      \csname LTb\endcsname
      \put(2658,250){\makebox(0,0){\strut{}\scriptsize 0}}%
      \csname LTb\endcsname
      \put(3063,250){\makebox(0,0){\strut{}\scriptsize 0.2}}%
      \csname LTb\endcsname
      \put(3467,250){\makebox(0,0){\strut{}\scriptsize 0.4}}%
      \csname LTb\endcsname
      \put(3872,250){\makebox(0,0){\strut{}\scriptsize 0.6}}%
      \csname LTb\endcsname
      \put(4276,250){\makebox(0,0){\strut{}\scriptsize 0.8}}%
      \csname LTb\endcsname
      \put(4681,250){\makebox(0,0){\strut{}\scriptsize 1}}%
    }%
    \gplgaddtomacro\gplfronttext{%
      \csname LTb\endcsname
      \put(3669,1952){\makebox(0,0){\strut{}\footnotesize fitted model simulation}}%
    }%
    \gplgaddtomacro\gplbacktext{%
      \csname LTb\endcsname
      \put(4946,250){\makebox(0,0){\strut{}\scriptsize 0}}%
      \csname LTb\endcsname
      \put(5350,250){\makebox(0,0){\strut{}\scriptsize 0.2}}%
      \csname LTb\endcsname
      \put(5755,250){\makebox(0,0){\strut{}\scriptsize 0.4}}%
      \csname LTb\endcsname
      \put(6159,250){\makebox(0,0){\strut{}\scriptsize 0.6}}%
      \csname LTb\endcsname
      \put(6564,250){\makebox(0,0){\strut{}\scriptsize 0.8}}%
      \csname LTb\endcsname
      \put(6968,250){\makebox(0,0){\strut{}\scriptsize 1}}%
    }%
    \gplgaddtomacro\gplfronttext{%
      \csname LTb\endcsname
      \put(7225,394){\makebox(0,0)[l]{\strut{}\tiny -3}}%
      \csname LTb\endcsname
      \put(7225,623){\makebox(0,0)[l]{\strut{}\tiny -2}}%
      \csname LTb\endcsname
      \put(7225,853){\makebox(0,0)[l]{\strut{}\tiny -1}}%
      \csname LTb\endcsname
      \put(7225,1083){\makebox(0,0)[l]{\strut{}\tiny 0}}%
      \csname LTb\endcsname
      \put(7225,1313){\makebox(0,0)[l]{\strut{}\tiny 1}}%
      \csname LTb\endcsname
      \put(7225,1543){\makebox(0,0)[l]{\strut{}\tiny 2}}%
      \csname LTb\endcsname
      \put(7225,1772){\makebox(0,0)[l]{\strut{}\tiny 3}}%
      \csname LTb\endcsname
      \put(5957,1952){\makebox(0,0){\strut{}\footnotesize fitted model simulation}}%
    }%
    \gplbacktext
    \put(0,0){\includegraphics[width={368.00bp},height={198.00bp}]{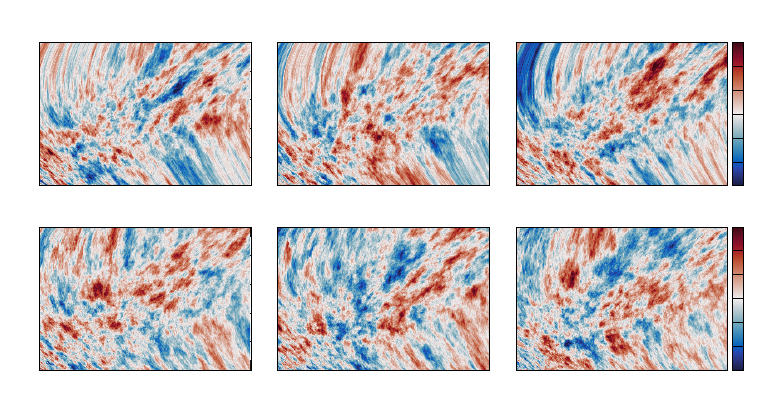}}%
    \gplfronttext
  \end{picture}%
\endgroup

    \caption{A comparison of simulations from the true nonstationary GP model
    given by \cref{eq:cov_ns} (top row) and simulations from the fitted
    nonparametric model (bottom row), using $m=200$ replicates from the true
    model, a random ordering, and $k=10$ nearest neighbors for each column of
    $\bT$.}
  \label{fig:simstudy}
\end{figure}

In this section, we study the ability of this estimator to recover the structure
of a highly nonstationary GP $Z_{\text{ns}}(\bx)$ with covariance
function given by
\begin{equation} \label{eq:cov_ns}
  K_{\text{ns}}(\bx, \bx') = \exp\left(-10 \left\|
  \bmsym{\varphi}(\bm{A}\bx) - \bmsym{\varphi}(\bm{A}\bx') \right\| \right),
\end{equation}
where $\bm{A}$ is a two-dimensional rotation matrix by $\pi/4$, and the non-linear
spatial deformation $\bmsym{\varphi}(\bm{v})$ for a coordinate vector $\bm{v} = (v_1,
v_2)\T$ is defined as $\bmsym{\varphi}(\bm{v}) = [\log(1 + 50v_1) + \log(1 + 50(v_2 +
1)), \, (1 + \exp(-8v_2))^{-1}]\T$.  Sample paths from this process on a $256
\times 256$ lattice on $[0,1]^2$ are shown in the top row of
\cref{fig:simstudy}. Several strong process features are exhibited in different
portions of the domain: in some cases, dependence is extremely strong along
nonlinear slices of the domain, and the anisotropy of the process changes
dramatically along and across the main diagonal. While particularly
pathological, features like shifting local anisotropy are ubiquitous in physical
and remote sensing data in the environmental sciences, for example the popular
local anisotropy model of \cite{paciorek2006}. The bottom row of
\cref{fig:simstudy} shows the results of simulating $m=200$ replicates from this
true model on the aforementioned lattice and fitting the above nonparametric
model using a random ordering and $k=10$ nearest-neighbor conditioning points in
each column of $\bT$.

As \cref{fig:simstudy} demonstrates, the fully nonparametric estimator
captures a great deal of the local structure very well: the rotating major axis
of anisotropy across the diagonal of the domain is very well-represented, and
the dramatic warpings in the off-diagonal corners are captured to a large degree
(although the near-singularity in the true warping function is a very non-local
feature that a sparse precision matrix of any form would be unlikely to capture
in full detail). A generic nonstationary covariance function approach that
captured so many local features would require tens if not hundreds of
parameters, and would yield a problematically non-convex likelihood surface that
would be very difficult to navigate---not to mention the extreme cost of
differentiating such a model. 

\begin{figure}[!ht]
  \centering
\begingroup
  \makeatletter
  \providecommand\color[2][]{%
    \GenericError{(gnuplot) \space\space\space\@spaces}{%
      Package color not loaded in conjunction with
      terminal option `colourtext'%
    }{See the gnuplot documentation for explanation.%
    }{Either use 'blacktext' in gnuplot or load the package
      color.sty in LaTeX.}%
    \renewcommand\color[2][]{}%
  }%
  \providecommand\includegraphics[2][]{%
    \GenericError{(gnuplot) \space\space\space\@spaces}{%
      Package graphicx or graphics not loaded%
    }{See the gnuplot documentation for explanation.%
    }{The gnuplot epslatex terminal needs graphicx.sty or graphics.sty.}%
    \renewcommand\includegraphics[2][]{}%
  }%
  \providecommand\rotatebox[2]{#2}%
  \@ifundefined{ifGPcolor}{%
    \newif\ifGPcolor
    \GPcolortrue
  }{}%
  \@ifundefined{ifGPblacktext}{%
    \newif\ifGPblacktext
    \GPblacktexttrue
  }{}%
  \let\gplgaddtomacro\g@addto@macro
  \gdef\gplbacktext{}%
  \gdef\gplfronttext{}%
  \makeatother
  \ifGPblacktext
    \def\colorrgb#1{}%
    \def\colorgray#1{}%
  \else
    \ifGPcolor
      \def\colorrgb#1{\color[rgb]{#1}}%
      \def\colorgray#1{\color[gray]{#1}}%
      \expandafter\def\csname LTw\endcsname{\color{white}}%
      \expandafter\def\csname LTb\endcsname{\color{black}}%
      \expandafter\def\csname LTa\endcsname{\color{black}}%
      \expandafter\def\csname LT0\endcsname{\color[rgb]{1,0,0}}%
      \expandafter\def\csname LT1\endcsname{\color[rgb]{0,1,0}}%
      \expandafter\def\csname LT2\endcsname{\color[rgb]{0,0,1}}%
      \expandafter\def\csname LT3\endcsname{\color[rgb]{1,0,1}}%
      \expandafter\def\csname LT4\endcsname{\color[rgb]{0,1,1}}%
      \expandafter\def\csname LT5\endcsname{\color[rgb]{1,1,0}}%
      \expandafter\def\csname LT6\endcsname{\color[rgb]{0,0,0}}%
      \expandafter\def\csname LT7\endcsname{\color[rgb]{1,0.3,0}}%
      \expandafter\def\csname LT8\endcsname{\color[rgb]{0.5,0.5,0.5}}%
    \else
      \def\colorrgb#1{\color{black}}%
      \def\colorgray#1{\color[gray]{#1}}%
      \expandafter\def\csname LTw\endcsname{\color{white}}%
      \expandafter\def\csname LTb\endcsname{\color{black}}%
      \expandafter\def\csname LTa\endcsname{\color{black}}%
      \expandafter\def\csname LT0\endcsname{\color{black}}%
      \expandafter\def\csname LT1\endcsname{\color{black}}%
      \expandafter\def\csname LT2\endcsname{\color{black}}%
      \expandafter\def\csname LT3\endcsname{\color{black}}%
      \expandafter\def\csname LT4\endcsname{\color{black}}%
      \expandafter\def\csname LT5\endcsname{\color{black}}%
      \expandafter\def\csname LT6\endcsname{\color{black}}%
      \expandafter\def\csname LT7\endcsname{\color{black}}%
      \expandafter\def\csname LT8\endcsname{\color{black}}%
    \fi
  \fi
    \setlength{\unitlength}{0.0500bp}%
    \ifx\gptboxheight\undefined%
      \newlength{\gptboxheight}%
      \newlength{\gptboxwidth}%
      \newsavebox{\gptboxtext}%
    \fi%
    \setlength{\fboxrule}{0.5pt}%
    \setlength{\fboxsep}{1pt}%
    \definecolor{tbcol}{rgb}{1,1,1}%
\begin{picture}(8500.00,3400.00)%
    \gplgaddtomacro\gplbacktext{%
      \csname LTb\endcsname
      \put(1171,338){\makebox(0,0)[r]{\strut{}\footnotesize $10^{-2}$}}%
      \csname LTb\endcsname
      \put(1171,997){\makebox(0,0)[r]{\strut{}\footnotesize $10^{-1}$}}%
      \csname LTb\endcsname
      \put(1171,1656){\makebox(0,0)[r]{\strut{}\footnotesize $10^{0}$}}%
      \csname LTb\endcsname
      \put(1171,2315){\makebox(0,0)[r]{\strut{}\footnotesize $10^{1}$}}%
      \csname LTb\endcsname
      \put(1272,98){\makebox(0,0){\strut{}\footnotesize $2^{10}$}}%
      \csname LTb\endcsname
      \put(1844,98){\makebox(0,0){\strut{}\footnotesize $2^{12}$}}%
      \csname LTb\endcsname
      \put(2416,98){\makebox(0,0){\strut{}\footnotesize $2^{14}$}}%
      \csname LTb\endcsname
      \put(2989,98){\makebox(0,0){\strut{}\footnotesize $2^{16}$}}%
      \csname LTb\endcsname
      \put(3561,98){\makebox(0,0){\strut{}\footnotesize $2^{18}$}}%
      \csname LTb\endcsname
      \put(4133,98){\makebox(0,0){\strut{}\footnotesize $2^{20}$}}%
    }%
    \gplgaddtomacro\gplfronttext{%
      \csname LTb\endcsname
      \put(1976,2657){\makebox(0,0)[r]{\strut{}\footnotesize $k=5$}}%
      \csname LTb\endcsname
      \put(1976,2417){\makebox(0,0)[r]{\strut{}\footnotesize $k=10$}}%
      \csname LTb\endcsname
      \put(1976,2177){\makebox(0,0)[r]{\strut{}\footnotesize $k=20$}}%
      \csname LTb\endcsname
      \put(1976,1938){\makebox(0,0)[r]{\strut{}\footnotesize $\bO(n)$}}%
      \csname LTb\endcsname
      \put(371,1605){\rotatebox{-270.00}{\makebox(0,0){\strut{}runtime (s)}}}%
      \csname LTb\endcsname
      \put(2702,2992){\makebox(0,0){\strut{}CPU}}%
    }%
    \gplgaddtomacro\gplbacktext{%
      \csname LTb\endcsname
      \put(5093,338){\makebox(0,0)[r]{\strut{}\footnotesize $10^{-2}$}}%
      \csname LTb\endcsname
      \put(5093,997){\makebox(0,0)[r]{\strut{}\footnotesize $10^{-1}$}}%
      \csname LTb\endcsname
      \put(5093,1656){\makebox(0,0)[r]{\strut{}\footnotesize $10^{0}$}}%
      \csname LTb\endcsname
      \put(5093,2315){\makebox(0,0)[r]{\strut{}\footnotesize $10^{1}$}}%
      \csname LTb\endcsname
      \put(5193,98){\makebox(0,0){\strut{}\footnotesize $2^{10}$}}%
      \csname LTb\endcsname
      \put(5766,98){\makebox(0,0){\strut{}\footnotesize $2^{12}$}}%
      \csname LTb\endcsname
      \put(6338,98){\makebox(0,0){\strut{}\footnotesize $2^{14}$}}%
      \csname LTb\endcsname
      \put(6911,98){\makebox(0,0){\strut{}\footnotesize $2^{16}$}}%
      \csname LTb\endcsname
      \put(7483,98){\makebox(0,0){\strut{}\footnotesize $2^{18}$}}%
      \csname LTb\endcsname
      \put(8055,98){\makebox(0,0){\strut{}\footnotesize $2^{20}$}}%
    }%
    \gplgaddtomacro\gplfronttext{%
      \csname LTb\endcsname
      \put(6624,2992){\makebox(0,0){\strut{}GPU}}%
    }%
    \gplbacktext
    \put(0,0){\includegraphics[width={425.00bp},height={170.00bp}]{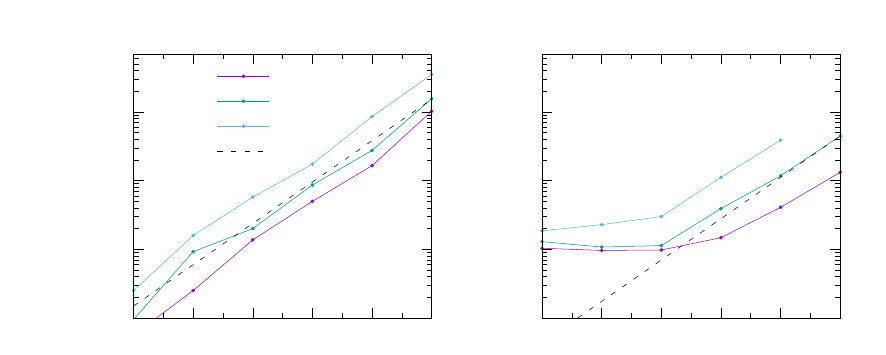}}%
    \gplfronttext
  \end{picture}%
\endgroup

    \caption{Empirical runtime costs of fitting the nonparametric model to data
    simulated using the true covariance function \cref{eq:cov_ns} with $n$ data
    points, $m=200$ replicates, and $k=\{5,10,20\}$ conditioning points per
    column of $\bT$. For the GPU runtime, the $n=2^{20}$, $k=20$ point is
    excluded due to RAM limitations.}
  \label{fig:runtime}
\end{figure}

\Cref{fig:runtime} shows the runtime cost of estimating the nonparametric $\bT$
from data simulated using this highly nonstationary model for a growing data
size $n$ and number of conditioning points $k$ on both CPU (AMD Ryzen 9950X),
which is run on a single thread except for BLAS and LAPACK elements that use as
many threads as physical cores, and on GPU (AMD Radeon Pro VII). At the largest
data size of $n=2^{20}$, the estimator takes approximately $10$ seconds to
compute with $k=5$ and $35$ seconds to compute with $k=20$ on a CPU. On a GPU, the
runtimes are sufficiently low until $n=2^{18}$ that the communication overhead
cost is dominant, with a runtime of just $4.4$ seconds with $n=2^{20}$ points
and $k=10$ conditioning points. Due to RAM limitations of the Radeon Pro VII
(16G), the last runtime for $n=2^{20}$ and $k=20$ was not possible to measure.
But based on extrapolation using the linear growth in $n$, the expected runtime
is approximately $15$ seconds.

\subsection{Application to computational fluid dynamics} \label{sec:cfd}

While the previous section demonstrates the method's ability to capture dramatic
nonstationarity in synthetic data, we now give a demonstration with a more
grounded application in computational fluid dynamics (CFD). Vinuesa et al.
\cite{vinuesa2015} describe a high-fidelity direct numerical simulation (DNS) of
fluid flow around a wall-mounted square cylinder in both laminar and turbulent
inflow regimes.  Datasets resulting from such a simulation offer several
challenges. First, to resolve local features, one must solve the Navier--Stokes
equations on high-order tensor-product quadrature rules. In this example, we use
data extracted from the center plane ($z=0$), and such a slice gives
measurements at a tensor product of a $y$-direction rule of $221$ nodes and an
$x$-direction rule of $441$ nodes, meaning that at each time snapshot one has $n
\approx 97\,000$ measurements.  As already indicated, these measurements are
\emph{not} made at regular step sizes in the $x$ or $y$ direction, and thus
cannot be treated as belonging to a lattice process. Second, and equally
importantly, these measurements naturally correspond to a nonstationary process,
as average streamwise velocity will naturally depend on, e.g., proximity to the
cylinder or other boundaries.  Importantly, these physical features of the
process will impact marginal properties, like variances, as well as more global
properties, like dominant eigenvectors.

\begin{figure}[!ht]
  \centering
  \makeatletter
\begingroup
  \makeatletter
  \providecommand\color[2][]{%
    \GenericError{(gnuplot) \space\space\space\@spaces}{%
      Package color not loaded in conjunction with
      terminal option `colourtext'%
    }{See the gnuplot documentation for explanation.%
    }{Either use 'blacktext' in gnuplot or load the package
      color.sty in LaTeX.}%
    \renewcommand\color[2][]{}%
  }%
  \providecommand\includegraphics[2][]{%
    \GenericError{(gnuplot) \space\space\space\@spaces}{%
      Package graphicx or graphics not loaded%
    }{See the gnuplot documentation for explanation.%
    }{The gnuplot epslatex terminal needs graphicx.sty or graphics.sty.}%
    \renewcommand\includegraphics[2][]{}%
  }%
  \providecommand\rotatebox[2]{#2}%
  \@ifundefined{ifGPcolor}{%
    \newif\ifGPcolor
    \GPcolortrue
  }{}%
  \@ifundefined{ifGPblacktext}{%
    \newif\ifGPblacktext
    \GPblacktexttrue
  }{}%
  \let\gplgaddtomacro\g@addto@macro
  \gdef\gplbacktext{}%
  \gdef\gplfronttext{}%
  \makeatother
  \ifGPblacktext
    \def\colorrgb#1{}%
    \def\colorgray#1{}%
  \else
    \ifGPcolor
      \def\colorrgb#1{\color[rgb]{#1}}%
      \def\colorgray#1{\color[gray]{#1}}%
      \expandafter\def\csname LTw\endcsname{\color{white}}%
      \expandafter\def\csname LTb\endcsname{\color{black}}%
      \expandafter\def\csname LTa\endcsname{\color{black}}%
      \expandafter\def\csname LT0\endcsname{\color[rgb]{1,0,0}}%
      \expandafter\def\csname LT1\endcsname{\color[rgb]{0,1,0}}%
      \expandafter\def\csname LT2\endcsname{\color[rgb]{0,0,1}}%
      \expandafter\def\csname LT3\endcsname{\color[rgb]{1,0,1}}%
      \expandafter\def\csname LT4\endcsname{\color[rgb]{0,1,1}}%
      \expandafter\def\csname LT5\endcsname{\color[rgb]{1,1,0}}%
      \expandafter\def\csname LT6\endcsname{\color[rgb]{0,0,0}}%
      \expandafter\def\csname LT7\endcsname{\color[rgb]{1,0.3,0}}%
      \expandafter\def\csname LT8\endcsname{\color[rgb]{0.5,0.5,0.5}}%
    \else
      \def\colorrgb#1{\color{black}}%
      \def\colorgray#1{\color[gray]{#1}}%
      \expandafter\def\csname LTw\endcsname{\color{white}}%
      \expandafter\def\csname LTb\endcsname{\color{black}}%
      \expandafter\def\csname LTa\endcsname{\color{black}}%
      \expandafter\def\csname LT0\endcsname{\color{black}}%
      \expandafter\def\csname LT1\endcsname{\color{black}}%
      \expandafter\def\csname LT2\endcsname{\color{black}}%
      \expandafter\def\csname LT3\endcsname{\color{black}}%
      \expandafter\def\csname LT4\endcsname{\color{black}}%
      \expandafter\def\csname LT5\endcsname{\color{black}}%
      \expandafter\def\csname LT6\endcsname{\color{black}}%
      \expandafter\def\csname LT7\endcsname{\color{black}}%
      \expandafter\def\csname LT8\endcsname{\color{black}}%
    \fi
  \fi
    \setlength{\unitlength}{0.0500bp}%
    \ifx\gptboxheight\undefined%
      \newlength{\gptboxheight}%
      \newlength{\gptboxwidth}%
      \newsavebox{\gptboxtext}%
    \fi%
    \setlength{\fboxrule}{0.5pt}%
    \setlength{\fboxsep}{1pt}%
    \definecolor{tbcol}{rgb}{1,1,1}%
\begin{picture}(9060.00,2820.00)%
    \gplgaddtomacro\gplbacktext{%
      \csname LTb\endcsname
      \put(577,622){\makebox(0,0)[r]{\strut{}\scriptsize 0.5}}%
      \csname LTb\endcsname
      \put(577,873){\makebox(0,0)[r]{\strut{}\scriptsize 1}}%
      \csname LTb\endcsname
      \put(577,1124){\makebox(0,0)[r]{\strut{}\scriptsize 1.5}}%
      \csname LTb\endcsname
      \put(577,1374){\makebox(0,0)[r]{\strut{}\scriptsize 2}}%
      \csname LTb\endcsname
      \put(577,1625){\makebox(0,0)[r]{\strut{}\scriptsize 2.5}}%
      \csname LTb\endcsname
      \put(577,1876){\makebox(0,0)[r]{\strut{}\scriptsize 3}}%
      \csname LTb\endcsname
      \put(577,2127){\makebox(0,0)[r]{\strut{}\scriptsize 3.5}}%
      \csname LTb\endcsname
      \put(577,2377){\makebox(0,0)[r]{\strut{}\scriptsize 4}}%
      \csname LTb\endcsname
      \put(877,300){\makebox(0,0){\strut{}\scriptsize 2}}%
      \csname LTb\endcsname
      \put(1141,300){\makebox(0,0){\strut{}\scriptsize 4}}%
      \csname LTb\endcsname
      \put(1405,300){\makebox(0,0){\strut{}\scriptsize 6}}%
      \csname LTb\endcsname
      \put(1669,300){\makebox(0,0){\strut{}\scriptsize 8}}%
      \csname LTb\endcsname
      \put(1933,300){\makebox(0,0){\strut{}\scriptsize 10}}%
      \csname LTb\endcsname
      \put(2197,300){\makebox(0,0){\strut{}\scriptsize 12}}%
      \csname LTb\endcsname
      \put(2461,300){\makebox(0,0){\strut{}\scriptsize 14}}%
      \csname LTb\endcsname
      \put(2725,300){\makebox(0,0){\strut{}\scriptsize 16}}%
    }%
    \gplgaddtomacro\gplfronttext{%
      \csname LTb\endcsname
      \put(180,1399){\rotatebox{-270.00}{\makebox(0,0){\strut{}\small y (unitless)}}}%
      \csname LTb\endcsname
      \put(1702,60){\makebox(0,0){\strut{}\small x (unitless)}}%
      \csname LTb\endcsname
      \put(1702,2559){\makebox(0,0){\strut{}\small Data snapshot (laminar)}}%
    }%
    \gplgaddtomacro\gplbacktext{%
      \csname LTb\endcsname
      \put(3394,622){\makebox(0,0)[r]{\strut{}\scriptsize 0.5}}%
      \csname LTb\endcsname
      \put(3394,873){\makebox(0,0)[r]{\strut{}\scriptsize 1}}%
      \csname LTb\endcsname
      \put(3394,1124){\makebox(0,0)[r]{\strut{}\scriptsize 1.5}}%
      \csname LTb\endcsname
      \put(3394,1374){\makebox(0,0)[r]{\strut{}\scriptsize 2}}%
      \csname LTb\endcsname
      \put(3394,1625){\makebox(0,0)[r]{\strut{}\scriptsize 2.5}}%
      \csname LTb\endcsname
      \put(3394,1876){\makebox(0,0)[r]{\strut{}\scriptsize 3}}%
      \csname LTb\endcsname
      \put(3394,2127){\makebox(0,0)[r]{\strut{}\scriptsize 3.5}}%
      \csname LTb\endcsname
      \put(3394,2377){\makebox(0,0)[r]{\strut{}\scriptsize 4}}%
      \csname LTb\endcsname
      \put(3694,300){\makebox(0,0){\strut{}\scriptsize 2}}%
      \csname LTb\endcsname
      \put(3958,300){\makebox(0,0){\strut{}\scriptsize 4}}%
      \csname LTb\endcsname
      \put(4222,300){\makebox(0,0){\strut{}\scriptsize 6}}%
      \csname LTb\endcsname
      \put(4486,300){\makebox(0,0){\strut{}\scriptsize 8}}%
      \csname LTb\endcsname
      \put(4750,300){\makebox(0,0){\strut{}\scriptsize 10}}%
      \csname LTb\endcsname
      \put(5015,300){\makebox(0,0){\strut{}\scriptsize 12}}%
      \csname LTb\endcsname
      \put(5279,300){\makebox(0,0){\strut{}\scriptsize 14}}%
      \csname LTb\endcsname
      \put(5543,300){\makebox(0,0){\strut{}\scriptsize 16}}%
    }%
    \gplgaddtomacro\gplfronttext{%
      \csname LTb\endcsname
      \put(4519,60){\makebox(0,0){\strut{}\small x (unitless)}}%
      \csname LTb\endcsname
      \put(4519,2559){\makebox(0,0){\strut{}\small Nonparametric simulation}}%
    }%
    \gplgaddtomacro\gplbacktext{%
      \csname LTb\endcsname
      \put(6212,622){\makebox(0,0)[r]{\strut{}\scriptsize 0.5}}%
      \csname LTb\endcsname
      \put(6212,873){\makebox(0,0)[r]{\strut{}\scriptsize 1}}%
      \csname LTb\endcsname
      \put(6212,1124){\makebox(0,0)[r]{\strut{}\scriptsize 1.5}}%
      \csname LTb\endcsname
      \put(6212,1374){\makebox(0,0)[r]{\strut{}\scriptsize 2}}%
      \csname LTb\endcsname
      \put(6212,1625){\makebox(0,0)[r]{\strut{}\scriptsize 2.5}}%
      \csname LTb\endcsname
      \put(6212,1876){\makebox(0,0)[r]{\strut{}\scriptsize 3}}%
      \csname LTb\endcsname
      \put(6212,2127){\makebox(0,0)[r]{\strut{}\scriptsize 3.5}}%
      \csname LTb\endcsname
      \put(6212,2377){\makebox(0,0)[r]{\strut{}\scriptsize 4}}%
      \csname LTb\endcsname
      \put(6512,300){\makebox(0,0){\strut{}\scriptsize 2}}%
      \csname LTb\endcsname
      \put(6776,300){\makebox(0,0){\strut{}\scriptsize 4}}%
      \csname LTb\endcsname
      \put(7040,300){\makebox(0,0){\strut{}\scriptsize 6}}%
      \csname LTb\endcsname
      \put(7304,300){\makebox(0,0){\strut{}\scriptsize 8}}%
      \csname LTb\endcsname
      \put(7568,300){\makebox(0,0){\strut{}\scriptsize 10}}%
      \csname LTb\endcsname
      \put(7832,300){\makebox(0,0){\strut{}\scriptsize 12}}%
      \csname LTb\endcsname
      \put(8096,300){\makebox(0,0){\strut{}\scriptsize 14}}%
      \csname LTb\endcsname
      \put(8360,300){\makebox(0,0){\strut{}\scriptsize 16}}%
    }%
    \gplgaddtomacro\gplfronttext{%
      \csname LTb\endcsname
      \put(7337,60){\makebox(0,0){\strut{}\small x (unitless)}}%
      \csname LTb\endcsname
      \put(8616,559){\makebox(0,0)[l]{\strut{}\scriptsize -1.5}}%
      \csname LTb\endcsname
      \put(8616,839){\makebox(0,0)[l]{\strut{}\scriptsize -1}}%
      \csname LTb\endcsname
      \put(8616,1119){\makebox(0,0)[l]{\strut{}\scriptsize -0.5}}%
      \csname LTb\endcsname
      \put(8616,1399){\makebox(0,0)[l]{\strut{}\scriptsize 0}}%
      \csname LTb\endcsname
      \put(8616,1679){\makebox(0,0)[l]{\strut{}\scriptsize 0.5}}%
      \csname LTb\endcsname
      \put(8616,1959){\makebox(0,0)[l]{\strut{}\scriptsize 1}}%
      \csname LTb\endcsname
      \put(8616,2239){\makebox(0,0)[l]{\strut{}\scriptsize 1.5}}%
      \csname LTb\endcsname
      \put(7337,2559){\makebox(0,0){\strut{}\small Mat\'ern simulation}}%
    }%
    \gplbacktext
    \put(0,0){\includegraphics[width={453.00bp},height={141.00bp}]{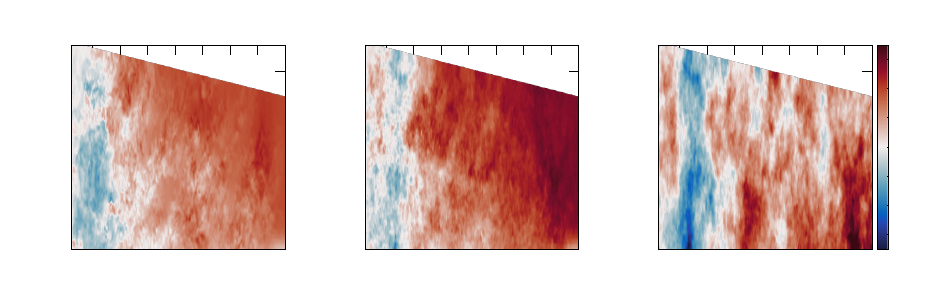}}%
    \gplfronttext
  \end{picture}%
\endgroup

  \makeatother
  \caption{Left: a sample snapshot of the laminar inflow condition data used to
  fit both models. Center and right: coupled simulations (using the same white
  noise forcing) from the nonparametric and Mat\'ern fitted models for the data.}
  \label{fig:cfd}
\end{figure}

Studying the marginal spatial dependence structure in DNS simulations is
necessary for several applications, such as characterizing temporal steady-state
behavior \cite{lumley1967,berkooz1993} or validating the accuracy of the
simulation by matching model-implied characteristics to theoretically known
behavior (see \cite{vinuesa2015,geoga2018} for more discussion). Motivated by
these needs, we investigate the efficacy of this nonparametric estimator by
fitting a standard Mat\'ern model alongside our estimator to this data using
precisely the same specification for a Vecchia approximation (in particular, the
same ordering and conditioning set design).  After thinning the time snapshots
by a factor of ten (leaving $31$ replicates in the case of laminar inflow and
$72$ for turbulent inflow) and standardizing each snapshot so as to better
isolate the ability of a model to capture dependence structure, we fit the two
models to the remaining samples, using a mild ridge parameter of $\gamma =
10^{-2}$ for the off-diagonal entries of the nonparametric estimator. For both
Vecchia specifications, a random ordering is used.  \Cref{fig:cfd,tab:cfd}
summarize the estimation results.

\begin{table}[!ht]
  \centering
  \begin{tabular}{ccc}
    \hline
    Case & Mat\'ern & Nonparametric \\
    \hline
    laminar   & $-381054$ & $\mathbf{-425202}$ \\
    turbulent & $-384800$ & $\mathbf{-457240}$ \\
    \hline
  \end{tabular}
  \caption{Mean log-score values for holdout data with estimated sparse inverse
  symmetric factors of the DNS data of \cite{vinuesa2015} simulated in laminar
  and turbulent inflow conditions. The log-score is the negative log-likelihood,
  so lower is better; the better of the two models is shown in bold.}
  \label{tab:cfd}
\end{table}

In \cref{fig:cfd}, we see a time snapshot of the laminar inflow condition data
and a coupled simulation between the two fitted models in the sense that they
are simulated with the same white noise forcings as $\bT^{-\intercal} \bm{z}$,
for $\bm{z} \sim \mathcal{N}(\bm{0}, \bm{I})$. For a discussion of fluid flow
characteristics, we refer the reader to \cite{vinuesa2015,geoga2018} and
references within. In the language of Gaussian processes, however, we see
several clear features. First, we note that the spatial dependence is generally
higher at larger $x$ values than smaller ones.  Second, we note strong negative
spatial correlation between values at large and small $x$ values, a feature that
is consistent in all snapshots of the real data. The former feature is a form of
nonstationarity. And while the latter may be possible to capture in a stationary
model, it certainly is \emph{not} possible to capture in a stationary model with
a covariance function for which no parameters allow covariances to be negative.
This is not a limitation that is unique to the Mat\'ern model, as very few
covariance functions allow for flexible negative dependence. Both of these
features that are difficult for kernel methods with off-the-shelf covariance
functions to produce are naturally and automatically captured by the
nonparametric model.

Finally, \cref{tab:cfd} quantifies the advantages discussed in the previous
paragraph by showing the mean log-score, which is just the negative
log-likelihood, of offset snapshots that were held out from fitting in each
case. In both cases, the nonparametric model decisively outperforms the Mat\'ern
model. In this particular demonstration, a small number of conditioning points
($k=5$) were used to specify the Vecchia approximation. From preliminary
investigation, however, the advantage of the nonparametric model is retained
even when the Vecchia specification for the Mat\'ern model is allowed many more
nonzeros per column of $\bT$: the holdout score still favors the nonparametric
estimator with $k=5$ over the Mat\'ern model with $k=20$, even though the extra
conditioning points give the latter more opportunities to reduce conditional
variances. Offering more degrees of freedom to the nonparametric model, on the
other hand, degrades the holdout log-score even as the fitted log-likelihood
improves, serving as an important reminder that overfitting in high-parameter
regimes like this one is a risk to be mindful of. This is precisely the regime
in which the ridge-type penalty of \cref{sec:opt_reg} is intended to be used; we
have not tuned it here, and doing so systematically is left to future work.

\subsection{Application to ensemble Kalman filtering} \label{sec:enkf}

A natural application of the nonparametric precision estimator is in the
ensemble Kalman filter (EnKF) \cite{Evensen1994, enkf}, where the forecast error
covariance $\bS_f$ must be estimated from a small ensemble of size $N_e$. The
sample covariance is rank-deficient (rank at most $N_e - 1$), and the standard
remedy is \emph{covariance localization}: element-wise multiplication of the
sample covariance with a compactly supported correlation function such as that of
Gaspari and Cohn \cite{GaspariCohn1999}, with a hand-tuned localization radius
\cite{HoutekamerMitchell2001}. While effective in many settings, this approach
has a fundamental limitation: the localization function must be chosen a priori,
which imposes an assumed correlation structure that may be poorly matched to the
true dynamics.

The sparse precision estimator of \cref{sec:optdesign} offers
an alternative. Rather than regularizing the covariance, we work directly in
precision space. The ensemble deviations $\{ \bx_j - \bar{\bx} \}_{j=1}^{N_e}$
serve as the $m = N_e$ replicates for the estimator, which produces a sparse
factor $\bT$ such that $\bT\bT\T \approx \bS_f\inv$. This
enables the \emph{precision-form} Kalman update:
\begin{align}
    \bOm_a &= \bT\bT\T + \bm{G}\T \bm{R}\inv \bm{G}, \label{eq:prec_update} \\
    \bx_a &= \bOm_a\inv \left( \bT\bT\T \bx_f + \bm{G}\T \bm{R}\inv \by \right), \label{eq:mean_update}
\end{align}
where $\bm{G}$ is the observation operator (written $\bm{G}$ rather than the
customary $\bm{H}$ to avoid a clash with the Hessian of
\cref{sec:optdesign}), $\bm{R}$ is the observation error covariance,
and $\bOm_a = \bS_a\inv$ is the analysis precision. This formulation is
attractive because $\bT\bT\T$ is sparse by construction, and the sparsity
pattern is determined by the geometry of the state space (via the Vecchia
approximation DAG) rather than by a hand-tuned parameter. The correlation
structure is learned from the ensemble at each assimilation step.

One practical consideration is that data being on different scales means that
the optimization problem needs to be tuned in several ways. First, box
constraints on the diagonal values of $\bT$ representing inverse square-root
conditional variances need to be rescaled. Secondly, stopping criteria may be
affected. To try and control for these issues and make the estimator more robust
to scaling, we apply a global standardization and normalize each ensemble member
to zero mean and unit variance before fitting, then rescale the resulting
estimate for $\bT$ accordingly. All estimates of $\bT$ in this section are
computed with NonparametricVecchia.jl and the interior-point solver MadNLP (see
\cref{sec:optdesign}).

A second consideration concerns the implementation of the update
\eqref{eq:prec_update}--\eqref{eq:mean_update}. A na\"ive implementation forms
$\bT\bT\T$, $\bOm_a$, and the Cholesky factor of $\bOm_a$ as dense $n
\times n$ matrices, requiring $3n^2$ entries of storage at every assimilation
step, and growing quadratically in $n$. We instead keep every
object sparse: $\bT\bT\T$ is computed as a sparse product, $\bm{G}\T \bm{R}\inv \bm{G}$
is assembled from sparse factors, the sum $\bOm_a = \bT\bT\T + \bm{G}\T \bm{R}\inv
\bm{G}$ remains sparse, and a sparse Cholesky factorization (\textsc{SuiteSparse}
\textsc{Cholmod}) is used for the analysis solve.
Writing $\text{nnz}(\cdot)$ for the number of stored nonzero entries of a sparse
matrix---so that $\text{nnz}(\bT) = p$ in the notation of
\cref{sec:optdesign}, and $\text{nnz}(\bT) = n(k+1)$ for the
$k$-nearest-neighbor conditioning sets used here---the analysis step has working
memory
\begin{equation} \label{eq:vec_mem}
  M_{\text{Vecchia}} \;=\; \text{nnz}(\bOm_a) + \text{nnz}(\text{chol}\,\bOm_a)
  + 2nN_e + n_{\text{obs}} N_e + 2 n_{\text{obs}},
\end{equation}
measured in stored floating-point numbers and scaling as $\bO(n k)$ in the sparse
structures, where $k$ is the number of conditioning neighbors. The last three
terms are the dense ensemble and observation-space blocks, which have direct
counterparts in the standard filter.
\Cref{tab:mem_accounting} itemizes the objects that make up the two working
sets.

The standard EnKF analysis stores the taper $\bmsym{\rho}$ that localizes the
cross-covariance between state and observations, the tapered product
$\bmsym{\rho} \circ \bS_f$ itself, the localized observation-space
matrix $\bm{G} \bS_f \bm{G}\T + \bm{R}$ together with its Cholesky factor, and
the dense ensemble blocks. The Gaspari--Cohn taper is compactly supported, vanishing for separations
beyond twice the radius $r$, so each of the first four is stored sparsely, giving
\begin{equation} \label{eq:gc_mem}
  M_{\text{GC}} \;=\; 2\,\text{nnz}(\bmsym{\rho}) + \text{nnz}(\bmsym{\rho}_{\text{obs}})
  + \text{nnz}(\text{chol}) + (n + 3 n_{\text{obs}}) N_e,
\end{equation}
with $\text{nnz}(\bmsym{\rho}) \approx n_{\text{obs}} (4r + 1)$ on a one-dimensional
grid, where $r$ is measured in grid spacings, and $\bmsym{\rho}_{\text{obs}}$ is the
restriction of $\bmsym{\rho}$ to the observed rows. Every object in both methods is
stored sparsely, and the two have approximately the same memory complexity,
linear in $n$.

\begin{table}[!ht]
\centering
\caption{Working memory of a single analysis step, itemized by object. Entries
are counts of stored floating-point numbers, and the two sparse totals are the
sparse parts of \eqref{eq:vec_mem} and \eqref{eq:gc_mem}; the dense blocks are
common to both methods and are dominated by the ensemble itself. All observations
in this example are point observations, so $\bm{G}\T \bm{R}\inv \bm{G}$ is diagonal and
$\bOm_a$ inherits the sparsity of $\bT\bT\T$ exactly.}
\label{tab:mem_accounting}
\small
\begin{tabular}{llc}
\hline
& Object & Entries \\
\hline
\multirow{5}{*}{Vecchia}
 & $\bOm_a = \bT\bT\T + \bm{G}\T \bm{R}\inv \bm{G}$ & $\text{nnz}(\bOm_a)$ \\
 & $\text{chol}(\bOm_a)$                   & $\text{nnz}(\text{chol}\,\bOm_a)$ \\
 & ensemble, state space                   & $2 n N_e$ \\
 & observation-space temporaries           & $n_{\text{obs}} N_e + 2 n_{\text{obs}}$ \\
\cline{2-3}
 & sparse total & $\text{nnz}(\bOm_a) + \text{nnz}(\text{chol}\,\bOm_a)$ \\
\hline
\multirow{6}{*}{Gaspari--Cohn}
 & $\bmsym{\rho}$                          & $\text{nnz}(\bmsym{\rho})$ \\
 & $\bmsym{\rho} \circ \bS_f$              & $\text{nnz}(\bmsym{\rho})$ \\
 & $\bm{G} \bS_f \bm{G}\T + \bm{R}$, localized    & $\text{nnz}(\bmsym{\rho}_{\text{obs}})$ \\
 & $\text{chol}(\bm{G} \bS_f \bm{G}\T + \bm{R})$  & $\text{nnz}(\text{chol})$ \\
 & ensemble and temporaries        & $(n + 3 n_{\text{obs}}) N_e$ \\
\cline{2-3}
 & sparse total & $2\,\text{nnz}(\bmsym{\rho}) + \text{nnz}(\bmsym{\rho}_{\text{obs}}) + \text{nnz}(\text{chol})$ \\
\hline
\end{tabular}
\end{table}

We demonstrate this approach against the standard EnKF with Gaspari--Cohn
localization on a dynamical system whose correlation length is a tunable
parameter, so that the regime in which each method is preferable can be exhibited
directly rather than argued for. Since reporting each method at its own best
setting would compare them at different budgets, we equalize the working memory
of the two analyses and compare accuracy within that common budget.

\subsubsection{Nonlocal advection-diffusion} \label{sec:nonlocal}

We consider a problem designed to exhibit long-range
correlations that are poorly matched by compactly supported localization
functions. The dynamics are governed by a one-dimensional nonlocal
advection-diffusion equation on the periodic domain $[0, 2\pi]$:
\begin{equation} \label{eq:nonlocal}
    \frac{\partial u}{\partial t} + v \frac{\partial u}{\partial x}
    = \alpha \int_0^{2\pi} \!\! K_{\lambda_c}(x - y) \, u(y) \, dy
    - \beta u + \nu \frac{\partial^2 u}{\partial x^2} + F(x,t),
\end{equation}
where $K_{\lambda_c}(\zeta) = e^{-|\zeta|/\lambda_c}$ is an exponential kernel with
correlation length $\lambda_c$. The domain is discretized on $n \in \{256, 512,
1{,}024, 2{,}048\}$ grid points and integrated with a fourth-order
Runge--Kutta scheme. To maintain stability under the CFL condition as the grid
is refined, we scale $\Delta t = 0.5 \Delta x$ and the viscosity proportionally
with $\Delta x$; the remaining parameters are $v = 1$, $\alpha = 0.5$, $\beta =
0.1$, with stochastic forcing $F$ at low wavenumbers.

The experimental setup uses an ensemble of $N_e = 30$ members with observations
at $10\%$ of randomly selected grid points, corrupted by Gaussian noise with
standard deviation $\sigma_{\text{obs}} = 0.1$, and covariance inflation of
$1.02$. Data are assimilated over $60$ steps with a $15$-step spinup, and the
RMSE values reported below are averages over the post-spinup steps of a single
realization.

\paragraph{Matched memory} For each conditioning size $k$ we tune the localization
radius so that $M_{\text{GC}} = M_{\text{Vecchia}}$ in
\cref{eq:vec_mem,eq:gc_mem}, giving a radius $r^{*}$ that depends on
$k$, the ensemble size and the observation density, but not on $n$: both memory
models are linear in $n$, so the matching condition is scale free.
Two caveats on this matching are worth stating plainly. First, the totals
\cref{eq:vec_mem,eq:gc_mem} include the dense ensemble blocks, which are common
to both methods and dominate at the grid sizes considered---roughly $80\%$ of the
budget at $n = 256$. Equalizing the totals therefore does \emph{not} equalize the
sparse structures: at $n = 256$, $k = 3$ the taper is granted about $1.7$ times
the sparse working set of the Vecchia factor, so the accounting is conservative
with respect to the localization rather than favorable to it. Second, the
differences reported below are single-realization averages without error bars, so
ratios within a few percent of unity should be read as ties.

\begin{table}[!ht]
\centering
\caption{Accuracy and working memory on the nonlocal advection-diffusion problem
\cref{eq:nonlocal}. The localization radius $r^{*}$, in grid spacings, is tuned so
that the two methods have equal working memory for one analysis step---the totals
\cref{eq:vec_mem,eq:gc_mem}, itemized in \cref{tab:mem_accounting}, in kilobytes.
The budgets match
exactly but for $n = 256$, $k = 15$, where the periodic domain caps $r^{*}$ and
the localization receives $87\%$ of the Vecchia total. The ratio is GC RMSE over
Vecchia RMSE; $> 1$ means Vecchia is more accurate at equal memory. Observations
are $10\%$ of grid points; $N_e = 30$. Values are single-realization averages
over the post-spinup steps.}
\label{tab:nonlocal_rmse}
\setlength{\tabcolsep}{6pt}
\begin{tabular}{rrrrccc}
\hline
& & & & \multicolumn{2}{c}{RMSE} & \\
\cline{5-6}
$n$ & $k$ & memory (KB) & $r^{*}$ & Vecchia & GC & ratio \\
\hline
\multicolumn{7}{c}{\emph{Long-range correlations, $\lambda_c = 1.0$}} \\
\hline
\multirow{3}{*}{$256$} & 3 & 152 & 39.5 & $2.01{\times}10^{-3}$ & $3.13{\times}10^{-3}$ & \textbf{1.55} \\
 & 8 & 182 & 56.5 & $1.93{\times}10^{-3}$ & $3.10{\times}10^{-3}$ & \textbf{1.61} \\
 & 15 & 223 & 63.5 & $1.92{\times}10^{-3}$ & $3.10{\times}10^{-3}$ & \textbf{1.61} \\
\hline
\multirow{3}{*}{$512$} & 3 & 304 & 40.5 & $1.73{\times}10^{-3}$ & $1.96{\times}10^{-3}$ & \textbf{1.13} \\
 & 8 & 364 & 57.5 & $1.71{\times}10^{-3}$ & $1.94{\times}10^{-3}$ & \textbf{1.13} \\
 & 15 & 449 & 81.5 & $1.71{\times}10^{-3}$ & $1.91{\times}10^{-3}$ & \textbf{1.12} \\
\hline
\multirow{3}{*}{$1{,}024$} & 3 & 608 & 40.5 & $1.21{\times}10^{-3}$ & $1.39{\times}10^{-3}$ & \textbf{1.15} \\
 & 8 & 730 & 57.5 & $1.21{\times}10^{-3}$ & $1.41{\times}10^{-3}$ & \textbf{1.17} \\
 & 15 & 900 & 81.5 & $1.21{\times}10^{-3}$ & $1.44{\times}10^{-3}$ & \textbf{1.20} \\
\hline
\multirow{3}{*}{$2{,}048$} & 3 & 1{,}216 & 40.0 & $7.35{\times}10^{-4}$ & $7.73{\times}10^{-4}$ & \textbf{1.05} \\
 & 8 & 1{,}461 & 57.5 & $7.35{\times}10^{-4}$ & $7.83{\times}10^{-4}$ & \textbf{1.07} \\
 & 15 & 1{,}803 & 81.0 & $7.35{\times}10^{-4}$ & $7.93{\times}10^{-4}$ & \textbf{1.08} \\
\hline
\multicolumn{7}{c}{\emph{Medium-range correlations, $\lambda_c = 0.5$}} \\
\hline
\multirow{3}{*}{$256$} & 3 & 152 & 39.5 & $5.65{\times}10^{-2}$ & $2.49{\times}10^{-2}$ & 0.44 \\
 & 8 & 182 & 56.5 & $5.80{\times}10^{-2}$ & $2.22{\times}10^{-2}$ & 0.38 \\
 & 15 & 223 & 63.5 & $5.82{\times}10^{-2}$ & $2.09{\times}10^{-2}$ & 0.36 \\
\hline
\multirow{3}{*}{$512$} & 3 & 304 & 40.5 & $6.50{\times}10^{-3}$ & $5.75{\times}10^{-3}$ & 0.88 \\
 & 8 & 364 & 57.5 & $6.43{\times}10^{-3}$ & $5.64{\times}10^{-3}$ & 0.88 \\
 & 15 & 449 & 81.5 & $6.44{\times}10^{-3}$ & $5.51{\times}10^{-3}$ & 0.86 \\
\hline
\multirow{3}{*}{$1{,}024$} & 3 & 608 & 40.5 & $2.11{\times}10^{-3}$ & $2.29{\times}10^{-3}$ & \textbf{1.08} \\
 & 8 & 730 & 57.5 & $2.11{\times}10^{-3}$ & $2.31{\times}10^{-3}$ & \textbf{1.09} \\
 & 15 & 900 & 81.5 & $2.11{\times}10^{-3}$ & $2.33{\times}10^{-3}$ & \textbf{1.10} \\
\hline
\multirow{3}{*}{$2{,}048$} & 3 & 1{,}216 & 40.0 & $9.52{\times}10^{-4}$ & $9.91{\times}10^{-4}$ & \textbf{1.04} \\
 & 8 & 1{,}461 & 57.5 & $9.52{\times}10^{-4}$ & $1.00{\times}10^{-3}$ & \textbf{1.05} \\
 & 15 & 1{,}803 & 81.0 & $9.52{\times}10^{-4}$ & $1.01{\times}10^{-3}$ & \textbf{1.06} \\
\hline
\end{tabular}
\end{table}

The two correlation lengths separate cleanly. At $\lambda_c = 1.0$ the Vecchia
EnKF is the more accurate at every grid and every conditioning size, by a factor
of up to $1.61$ at $n = 256$. The margin does not close as the budget grows: at
$n = 1{,}024$ it widens with $k$, from $1.15$ to $1.20$, the wider radius that
the larger budget buys readmitting the spurious long-range sample correlation
that localization exists to suppress. At $n = 2{,}048$ the two methods are within
$8\%$ of each other, which on a single realization we regard as a tie.

At $\lambda_c = 0.5$ the ranking reverses, most sharply on the coarsest grid. At $n
= 256$ the post-spinup truth standard deviation is $5.5 \times 10^{-2}$, so an
analysis that ignored the observations entirely and returned the climatological
mean would incur that error. The Vecchia analysis error is $5.6$ to $5.8 \times
10^{-2}$: it does not improve on that no-skill baseline. The localization, at
$2.1$ to $2.5 \times 10^{-2}$, reduces the error by $54$ to $62\%$. We attribute
this to the estimation step rather than to the precision form. Because the
objective separates by column, the relevant sample-size ratio is $m/(k+1)$ per
column, not $N_e$ against the $n(k+1)$ entries of $\bT$: with $m = N_e = 30$ this
ratio falls from $7.5$ at $k=3$ to $1.9$ at $k=15$, approaching the $m = k+1$
threshold at which the blocks $\bm{H}_j$ become singular
(\cref{sec:optdesign}). The rows at $n = 256$ are consistent with this reading,
the error degrading monotonically in $k$ exactly where the ratio is smallest,
whereas at $\lambda_c = 1.0$ the additional neighbors still pay for themselves.
The gap closes quickly with
resolution: the two methods are within $16\%$ of each other at $n = 512$, and
from $n = 1{,}024$ onward the Vecchia filter is nominally ahead at both
correlation lengths, though by margins ($4$ to $10\%$) that a single realization
does not resolve. The defensible reading of \cref{tab:nonlocal_rmse} is
therefore that the precision form wins clearly when the correlations are long
relative to any admissible taper, loses clearly when the ensemble is too small
to estimate $\bT$ at all, and is otherwise comparable.

\section{Discussion} \label{sec:discussion}

We have introduced a nonparametric estimator for sparse precision matrices based
on Vecchia approximation with a convex optimization formulation. The estimator
produces a sparse inverse Cholesky factor $\bT$ with $\bO(nk)$ nonzeros at a
cost of $\bO(nmk^2 + nk^3)$ for the one-time assembly and factorization of the
block-diagonal Hessian plus $\bO(nk^2)$ per interior-point iteration, all linear
in $n$ at fixed $k$. The problem is convex in $\bT$, and the smooth
reparameterization we solve admits an efficient interior-point solution via
condensed KKT systems whose blocks are either diagonal or block diagonal with
$(k+1) \times (k+1)$ blocks. As demonstrated in several examples, this approach
for estimating $\bT$ is extremely fast and provides the expressiveness one
expects from a nonparametric estimator.

In many ways, however, this present form of the estimator is the beginning of a
larger project. Most importantly, the regularization mechanisms described
here---box constraints on the conditional variances and ridge-type penalties on
the entries of $\bT$, and thus on the Kriging weights---are effective enough for
small $k$, but for larger numbers of conditioning sets the risk of overfitting
is high. The CFD example of \cref{sec:cfd} shows this directly: the holdout
log-score degrades as degrees of freedom are added, and we have not attempted to
tune the ridge penalty against it. Alternative methods of regularization that do
not harm the scalability of the estimation may be extremely valuable.

The ensemble Kalman filtering results of \cref{sec:enkf} suggest a second line
of work. At equal working memory, the precision form is clearly preferable when
the true correlations are long relative to any admissible taper, and clearly
worse when the ensemble is too small to estimate $\bT$ reliably---a failure of
the estimation step rather than of the update. Characterizing that transition,
and in particular determining how the conditioning size $k$ must grow with the
correlation length to maintain a given accuracy, would make the tradeoff
predictable in advance rather than empirical. Larger and repeated experiments,
with uncertainty quantification across realizations, are needed to settle the
regimes where the two methods are close.

A third natural extension and improvement to this approach concerns GPUs. The
specialized KKT solver of \cref{sec:optdesign} already runs on NVIDIA and AMD
GPUs, with one batched Cholesky factorization of the blocks $\bm{H}_j$ and
$\bO(n)$ diagonal operations per interior-point iteration, and may offer very
significant speedups that would make the method even more computationally
economical than it presently is. A systematic GPU benchmark of the full
estimation pipeline is left for future work. The column-separable structure of
\cref{eq:problem}, which already allows the estimation to be carried out in
independent batches within a fixed memory budget, is well suited to such an
implementation.

\newpage

\renewcommand{\abstractname}{Acknowledgements}
\begin{abstract}
This material is based on work supported by the National Science Foundation
under Grant Number DMS-2610202 and the U.S. Department of Energy, Office of Science,
Office of Advanced Scientific Computing Research (ASCR) under
Contract DE-AC02-06CH11347.  Any opinions, findings, and conclusions or
recommendations expressed in this material are those of the author(s) and do not
necessarily reflect the views of the National Science Foundation or the Department of Energy.
\end{abstract}


\bibliographystyle{plain}
\bibliography{references}

\begin{flushright}
	\scriptsize \framebox{\parbox{3.4in}{Government License: The
			submitted manuscript has been created by UChicago Argonne,
			LLC, Operator of Argonne National Laboratory (``Argonne").
			Argonne, a U.S. Department of Energy Office of Science
			laboratory, is operated under Contract
			No. DE-AC02-06CH11357.  The U.S. Government retains for
			itself, and others acting on its behalf, a paid-up
			nonexclusive, irrevocable worldwide license in said
			article to reproduce, prepare derivative works, distribute
			copies to the public, and perform publicly and display
			publicly, by or on behalf of the Government. The Department of Energy will provide public access to these results of federally sponsored research in accordance with the DOE Public Access Plan. http://energy.gov/downloads/doe-public-access-plan. }}
	\normalsize
\end{flushright}

\end{document}